\documentclass[fleqn,usenatbib]{mnras}

\usepackage{newtxtext,newtxmath}

\newcommand{\Ha}{$\rm{H} \alpha$}
\newcommand{\Hb}{$\rm{H} \beta$}

\newcommand{\HI}{\hbox{{\rm H}\kern 0.1em{\sc i}}}
\newcommand{\Lya}{\hbox{{\rm Ly}\kern 0.1em$\alpha$}}
\newcommand{\Lyb}{\hbox{{\rm Ly}\kern 0.1em$\beta$}}
\newcommand{\MgII}{\hbox{{\rm Mg}\kern 0.1em{\sc ii}}}
\newcommand{\SiII}{\hbox{{\rm Si}\kern 0.1em{\sc ii}}}
\newcommand{\CIII}{\hbox{{\rm C}\kern 0.1em{\sc iii}}}
\newcommand{\CIV}{\hbox{{\rm C}\kern 0.1em{\sc iv}}}
\newcommand{\NII}{\hbox{{\rm N}\kern 0.1em{\sc ii}}}
\newcommand{\OVI}{\hbox{{\rm O}\kern 0.1em{\sc vi}}}
\newcommand{\OII}{\hbox{[{\rm O}\kern 0.1em{\sc ii}]}}
\newcommand{\OIII}{\hbox{[{\rm O}\kern 0.1em{\sc iii}]}}
\newcommand{\kms}{\hbox{~km~s$^{-1}$}}

\newcommand{\colden}{\ensuremath{\log (N(\HI)/{\rm cm}^{-2})}}
\newcommand{\fewcorot}{\hbox{$f_{\rm EWcorot}$}}
\usepackage[T1]{fontenc}

\DeclareRobustCommand{\VAN}[3]{#2}
\let\VANthebibliography\thebibliography
\def\thebibliography{\DeclareRobustCommand{\VAN}[3]{##3}\VANthebibliography}
\usepackage{pdflscape}
\usepackage{lscape}

\usepackage{graphicx}	
\usepackage{amsmath}	
\usepackage{multirow}
\usepackage{setspace}
\usepackage{bigdelim}
\usepackage{bigstrut}
\usepackage{floatflt}	
\usepackage{wrapfig}
\usepackage{subfig}
\usepackage{multicol}
\usepackage[utf8]{inputenc}
\usepackage[flushleft]{threeparttable}
\usepackage{xltabular}
\usepackage{caption}

\title[Galaxy--{\HI} CGM kinematics
]{Signatures of Gas Flows-I: Connecting the kinematics of the {\HI} circumgalactic medium to galaxy rotation}

\author[Nateghi et al.]
{Hasti Nateghi,$^{1,2}$\thanks{E-mail: hnateghi@swin.edu.au}
Glenn G. Kacprzak$^{1,2}$,
Nikole M. Nielsen$^{1,2,3}$, 
Michael T. Murphy$^{1}$,
\newauthor Christopher W. Churchill$^{4}$, Sowgat Muzahid$^{5}$, Sameer$^{6,7}$, 
Jane C. Charlton$^{6}$\\
$^{1}$Centre for Astrophysics and Supercomputing, Swinburne University of Technology, Hawthorn, Victoria 3122, Australia\\
$^{2}$ARC Centre of Excellence for All Sky Astrophysics in 3  Dimensions (ASTRO 3D), Australia\\
$^{3}$Homer L. Dodge Department of Physics and Astronomy, The University of Oklahoma, 440 W. Brooks St., Norman, OK 73019, USA\\
$^{4}$Department of Astronomy, New Mexico State University, Las Cruces, NM 88003, USA\\
$^{5}$Inter-University Centre for Astronomy and Astrophysics (IUCAA), Post Bag 4, Ganeshkhind, Pune 411 007, India\\
$^{6}$Department of Physics and Astronomy, The University of Notre Dame, Notre Dame, IN 46544, USA\\
$^{7}$Department of Astronomy and Astrophysics, The Pennsylvania State University, State College, PA 16801, USA\\
}

\date{Accepted 2024 July 26. Received 2024 July 18; in original form 2023 March 10}

\pubyear{2024}

\begin{document}
\label{firstpage}
\pagerange{\pageref{firstpage}--\pageref{lastpage}}
\maketitle

\begin{abstract}

The CGM hosts many physical processes with different kinematic signatures that affect galaxy evolution. We address the CGM--galaxy kinematic connection by quantifying the fraction of {\HI} that is aligned with galaxy rotation with the equivalent width co-rotation fraction, {\fewcorot}. Using 70 quasar sightlines having {\it HST}/COS {\HI} absorption (${12<\colden<20}$) within $5R_{\rm vir}$ of $z<0.6$ galaxies we find that {\fewcorot} increases with increasing {\HI} column density. {\fewcorot} is flat at $\sim0.6$ within $R_{\rm vir}$ and decreases beyond $R_{\rm vir}$ to {\fewcorot}$\sim0.35$. {\fewcorot} also has a flat distribution with azimuthal and inclination angles within $R_{\rm vir}$, but decreases by a factor of two outside of $R_{\rm vir}$ for minor axis gas and by a factor of two for edge-on galaxies. Inside $R_{\rm vir}$, co-rotation dominated {\HI} is located within $\sim 20$~deg of the major and minor axes.  We surprisingly find equal amounts of {\HI} absorption consistent with co-rotation along both major and minor axes within $R_{\rm vir}$. However, this co-rotation disappears along the minor axis beyond $R_{\rm vir}$, suggesting that if this gas is from outflows, then it is bound to galaxies. {\fewcorot} is constant over two decades of halo mass, with no decrease for log(M$_{\rm h}/M_{\odot})>12$ as expected from simulations. Our results suggest that co-rotating gas flows are best found by searching for higher column density gas within $R_{\rm vir}$ and near the major and minor axes.

\end{abstract}




\section{Introduction}


Gas accretion through the circumgalactic medium (CGM) plays a major role in the growth and evolution of galaxies.
Galaxies hierarchically form and evolve via gas flows onto them, which originates from the cosmic web, tidal streams, galaxy mergers, galactic winds, and fountains. Cosmological simulations predict that accretion occurs via two modes depending on whether the gas is shock heated or not \citep[e.g.,][]{keres2005, Dekel_Birnboim2006,Dekel2009, FaucherGiguere&keres2011,stewart2011a, stewart2011b,  stewart2013,vandeVoort2011,Hobb2015}. 
Focusing on the cold-mode of accretion, \citet{Danovich2015} used cosmological simulations and showed that the kinematics of galaxy disks are comparable to the spin of dark matter halos regardless of the gas and dark matter angular momentum histories. 
Their results are consistent with \citet{stewart2017} who tested five hydrodynamic codes and concluded that the ubiquitous presence of co-directional, co-planar filamentary accretion, with higher angular momentum than dark matter, can support the $\Lambda$CDM prediction in galaxy formation. However, observation of CGM gas accretion is difficult and challenging.

 A large number of studies have shown that the vast majority of low ionisation metal-line absorption exhibits co-directional and co-planar accretion kinematics \citep{Steidel2002, Glenn2010a, Glenn2011june, Bouche13,burchett2013, Jorgenson-wolf2014, bouche2016, Ho17,Rahmani2018, MartinCrystal2019, Lopez2020}. In these studies, it was noted that the majority or bulk of the absorption seems to align with the rotation direction of the host galaxy. \citet{Steidel2002} used a corotating thick disk model to confirm an extended rotating disk-like structure with some velocity lag is a plausible explanation for the {\MgII} kinematics detected in the galaxy halos. Furthermore, these signatures of co-rotation in {\MgII} absorption were investigated by \citet{Glenn2010a} and \citet{Ho17} who also found that the bulk of absorption is consistent with observed galaxy rotation. They inferred that the absorbing gas kinematics is either lagging in rotation or infalling.  
 However in these works, absorption systems were counted as either co-rotating or not, without quantifying how much gas was associated with co-rotation.

For the first time, {\OVI} halo--galaxy relative kinematics was examined by \citet{Glenn2019} who found that despite the {\MgII} absorption, major axis {\OVI} is not likely related to host galaxies' rotation. However, they could explain the kinematics of {\OVI} detected along the minor axis as outflows with small opening angles and they concluded that {\OVI} that originates from a diffuse high ionisation phase of CGM is likely not a good kinematic indicator for ongoing processes in the CGM. 

Some observations of {\MgII} absorption have shown a bimodal picture where the majority of CGM gas has been detected along the galaxies' major and minor axes \citep{Bordoloi2011, Bouche2012, Glenn2015decmorpho, Lan2018,Langan2023MegaflowIX}.  These results have inferred that this gas originates from accretion and outflows, respectively. This is further supported by kinematic studies that show signatures of accreting {\MgII} gas along the galaxy major axes \citep{Steidel2002,Ho17, Diamond-Stanic2016, Zabl19_megaflow2}, and outflowing along their minor axes \citep[e.g,][]{Bouche2012,Schroetter19_megaflow3}.  However, while {\OVI} is also distributed bimodally along the major and minor axes, kinematic studies of {\OVI} show no strong kinematic correlation or signatures of accretion or outflows \citep{Glenn2015decmorpho, Nikki2017,Glenn2019,MasonNg2019}. \citet{Glenn2019} also used simulations to suggest that although gas flows are present, they may be masked by a diffuse {\OVI} component. 

Various ions have been used to study the CGM that samples different gas densities and temperatures, however, {\HI} may bridge the gap between the low- and high-ionisation halos studied in previous works. It is well known that {\HI} tracks both the low and high ionisation CGM and can be associated with a variety of environments and gas densities like cosmic web filamentary inflows, galactic feedback, tidal stripping caused by mergers, and surrounding {\HI} clouds. 
So understanding how the {\HI} is kinematically coupled to the rotation of galaxy disks may provide new insights into ongoing gas processes.
Cosmological simulations have shown that the high column density {\HI} gas in the halo is mostly associated with gas flows in and out of the host galaxies \citep{Fumagalli2011,vandeVoort2012,suresh2019}. In this paper, we observationally test this scenario and examine whether the gas accretion/outflow is dependent on {\HI} column density.

The kinematic relation between {\Lya} absorption line and the host galaxy was initially studied by \cite{Barcon1995} who found consistency between the kinematics of stellar disks and the halo of two galaxies at $z=$0.075 and 0.09 and showed that the {\Lya} gas corotates with the inner disk of the galaxies. \cite{Cote2005} also studied the kinematics of nine {\HI} halos at large galactocentric distances and found an inconsistency between the lower column density {\Lya} absorption and disk rotation in three systems that can confirm the expectation of cosmic web origin of the gas. 

A recent study by \citet{French2020} showed that up to $59\% \pm 5\%$ of {\Lya} absorbers in their sample have consistent kinematics with their host galaxies. They also found an anti-correlation between the corotation fraction of {\HI} and its projected distance from the host galaxies as well as galaxies' luminosity and inclination angle.
In a step forward in methodology, \citet{French2020} decomposed their {\Lya} absorption into multiple components and counted each component separately in order to measure the co-rotation fraction of {\Lya} absorption. This better quantified how much gas in each absorption system is consistent with a co-rotation model. However, this approach only works for low column density, unsaturated absorption systems, with no complex velocity structure. When absorption systems have a complex velocity structure or are saturated, then results will be dependent on how many components one fits into the data and the assumptions being made, e.g., assume the {\HI} has the same velocity structure of the metal lines, or use the least amount of fitted components to achieve the best fit, etc. In this study, we have taken a new approach to quantify the amount of gas that has kinematics consistent with co-rotation, which relies on the data rather than user/model absorption decomposition.

Using the quasar absorption line technique, we quantify the kinematic connections between the CGM {\HI} gas and their host galaxies in 70 galaxy-CGM absorption pairs. The high resolution galaxy spectra obtained by the Echelle Spectrograph and Imager \citep[ESI,][]{Sheinis2002} on Keck~II provided us with the rotation curves for most galaxies. We also measured the {\HI} gas properties such as kinematics, equivalent widths, and column densities using the {\Lya} absorption lines detected in the background quasar spectra observed with the Cosmic Origins Spectrograph (COS) on the {\it Hubble Space Telescope} ({\it HST}). The quasar sightlines in our sample trace {\HI} gas within projected distances of $10\leq D \leq815$~kpc of galaxies over the redshift range of $z=0.002-0.55$. Here for the first time, we measure the fraction of {\HI} equivalent width in each system that could be kinematically coupled with the rotation of its host galaxy to avoid any fitting and model dependencies and to provide a better estimate of co-rotating gas around galaxies. 

The paper is organized as follows: 
In Section~\ref{sec:method} we describe the data and analysis. This included our new method for quantifying the co-rotation fraction ({\fewcorot}). In Section~\ref{sec:results} we present the results of how {\fewcorot} varies as a function of the {\HI} column density, impact parameter, virial radius normalised impact parameter, azimuthal angle, galaxy inclination angle, stellar and halo mass. In Section~\ref{sec:discussion} we discuss our results and present our concluding remarks in Section~\ref{sec:conslusion}.
Throughout we adopt an ${\rm H}_{\rm 0}=70$~\kms~Mpc$^{-1}$, $\Omega_{\rm M}=0.3$, $\Omega_{\Lambda}=0.7$ cosmology.


\begin{figure*}
    \centering
    \includegraphics[width=0.8\linewidth]{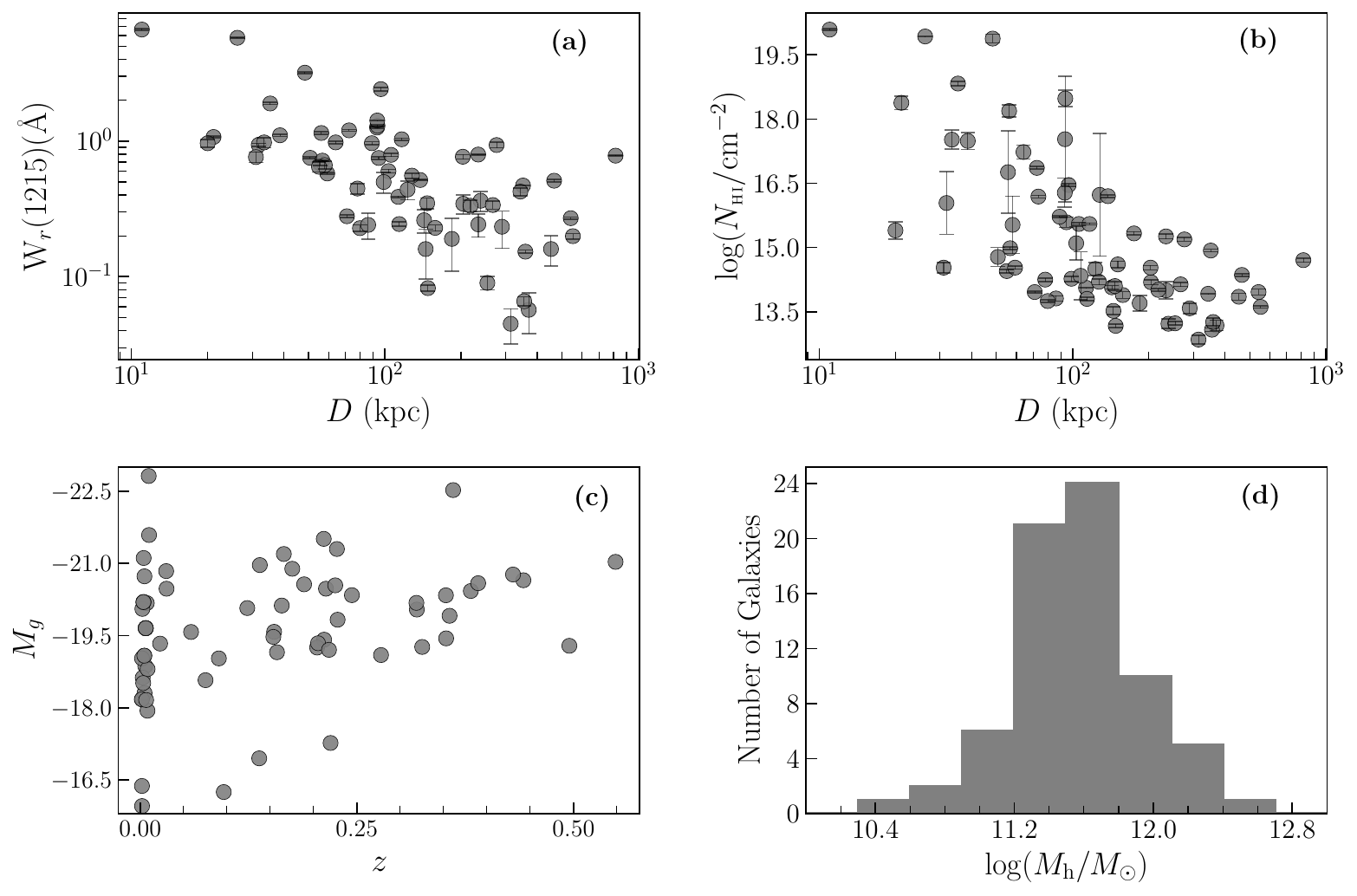}\hfill 
    \caption{(a) Rest-frame equivalent width of {\Lya} ($W_r(1215)$) as a function of impact parameter ($D$). We see a clear anti-correlation between absorption strength and impact parameter, which is consistent with the literature. (b) Distribution of {\HI} absorption column densities in our sample as a function of projected distance from galaxies. (c) The g-band absolute magnitude of galaxies as a function of redshift. (d) The halo mass distribution of the galaxies in our sample. }
    \label{fig:sample properties}
\end{figure*}

\section{Observations and Analysis}
\label{sec:method}

Our sample comprises 70 galaxy--{\HI} absorption pairs that span a redshift range of $z=0.002-0.55$ and within impact parameters of $D=10-815$~kpc. Every galaxy has {\Lya} absorption detected in {\it HST}/COS G130M or G160M quasar spectra with column densities ranging between $\log (N({\HI})/{\rm cm}^{-2})=12.6-20.9$. We have obtained the kinematics/rotation curves of 39 galaxies with Keck/ESI, which were observed as part of the Multiphase Galaxy Halos survey \citep[e.g.,][]{Glenn2015decmorpho, Nikki2017, pointon19, Hasti2021}. The remaining 23 galaxy kinematics were obtained from  \citet{French2020}. Four galaxies having multiple quasar sightlines (see Table~\ref{galaxies}), which result in a total of 62 galaxies for the entire sample and 70 galaxy-absorption pairs. 
The sample contains a mixture of galaxy-selected \citep{pointon19, French2020} and absorption-selected \citep{Tripp2008} absorber--galaxy pairs.

We focus on isolated galaxies in order to reduce any environmental effects, such as perturbations on the galaxy rotation curves or gas distributions due to interactions or major mergers which complicate correlations between galaxy and CGM kinematics \citep[e.g.,][]{Pointon2017,Nielsen2018_magiiVI,nikki2022,antonia2024}. For our higher redshift galaxies selected from \citet{pointon19}, they report that there are no major companions within 100~kpc and with velocity separations less than 500~\kms. For the low redshift galaxies selected from \citet{French2020}, they report that the galaxies are relatively isolated based on their likelihood criteria and within 3~$R_{\rm vir}$ of a background quasar. In both samples,  galaxies may still have nearby minor companions, which likely do not affect the kinematics of the larger galaxy.
Summaries of the {\HI} observations and galaxy sample are presented in  Table~\ref{QSOs} and Table~\ref{galaxies}, respectively. A summary of the sample is also presented in Fig.~\ref{fig:sample properties}. Fig.~\ref{fig:sample properties}(a) shows the rest-frame equivalent width of {\Lya} ($W_r(1215)$) and (b) shows the {\HI} column density (\colden) as a function of impact parameter ($D$). These panels show the strong anti-correlation between absorption strength and $D$. Fig.~\ref{fig:sample properties}(c) and (d) shows the absolute magnitude and halo mass distributions of our sample, respectively. We describe the details of the data for this sample in the following subsections.

\subsection{Galaxy morphologies}
\label{imaging}
In order to connect the CGM absorption to the host galaxies, we need the galaxy morphologies and the alignment of the quasar sightline relative to the galaxy disks (i.e., inclination and azimuthal angles).  
We obtained previously published inclination angles and azimuthal angles for 23 galaxies \citep{French2020}, which are listed in Table~\ref{galaxies}. We further obtained previously published morphologies/geometries for 28 galaxies \citep{Glenn2015decmorpho,Glenn2019,Glenn2019b_metal}, which were computed using GIM2D \citep{Simard2002} models of {\it HST} images with ACS, WFC3, or WFPC2 in the F702W, F814W, or F625W filters as listed in Table~\ref{QSOs}.  Here we add new models for 5 galaxies having {\it HST} images and 5 galaxies with Pan-STARRS \citep[hereafter PS;][]{chambers16} images, which are listed in Table~\ref{QSOs}. The orientation of the galaxies, such as their inclination ($i$) and azimuthal angle ($\Phi$), were modelled following the methods adopted from \citet{Glenn2011augustGIM2D} and \citet{Glenn2015decmorpho}. We fit two-component disc+bulge models using GIM2D \citep{Simard2002} to the {\it HST} and PS images using modelled point spread functions \citep[see][]{Glenn2015decmorpho}. The galaxy disk component is modelled with an exponential profile and the bulge component has a Sersic profile with $0.2<n <4.0$. 
The modelled inclination and azimuthal angles, and their errors, for all galaxies are listed in Table~\ref{galaxies}.
 We adopt the convention of the azimuthal angle ${{\Phi }}=0^\circ $ to be along the galaxy major axis and ${{\Phi }}=90^\circ $ to be along the galaxy minor axis.

Our galaxies have a full range of azimuthal and inclination angles (see Table~\ref{galaxies}). We test for potential effects of a biased distribution in azimuthal and inclination angle on our results by conducting a one-dimensional Kolmogorov-Smirnov (KS) test on $\Phi$ and $i$ distributions. Our analysis indicates that the azimuthal angle is consistent with the expected flat distribution for a random sample of galaxies at a significance level of $1.36\sigma$. We also find that the inclination angles are consistent with the expected sin($i$) distribution for a random sample of galaxies \citep{Law2009}, at the significance level of $1.84\sigma$. Although a preference towards edge-on galaxies is ideal for examining outflows and inflows, we have determined that our results remain unchanged within the errors reported here when we exclude galaxies with $i<30$~deg (14 galaxies in total). Thus, we include all galaxies in our analysis. 

\subsection{Galaxy photometries and masses} 
The behaviour of the CGM is dependent on galaxy mass and its properties vary with location within the virial radius of the halo \citep{chen2010may, Churchil13_selfsimilar, churchill2013_magii3, Tumlinson13, oppenheimer16, MasonNg2019}. 
Therefore we have computed the stellar masses for all galaxies using the rest-frame $g-r$ colour and $g$-band mass-to-light ratio ($M/L$) relation from \citet{Bell2003}. Galactic extinction corrections are not applied as they are on the order of the uncertainties in our method. Given the range of redshifts and galaxy angular extents in our sample, we have used a range of catalogues, such as PS, SDSS \citep{SDSS_York2000}, 
 and DESI Legacy surveys \citep{Dey2019_DESI}, to obtain the galaxy photometry and colours. 

For galaxies with $z>0.05$, we obtained $g$ and $r$ Kron magnitudes from PS. In cases where PS photometry was not available, we used either SDSS model magnitudes or {\it HST} photometry.  To apply the $M/L$ ratio relation, the rest-frame colours of galaxies are required. We applied $K$-corrections following the methods described by \citet{Nikki13a_magiicat1} to obtain rest-frame absolute $g$- and $r$-band magnitudes.
For the galaxies with no $g-r$ observed colour, we assumed an Sbc type, which represents the typical galaxy colour found for galaxies associated with CGM absorption. \citep{Steidel94,zibetti2007, Nikki13a_magiicat1,Glenn2015decmorpho}.

Galaxies with $z<0.05$ have larger angular extents, which typically results in underestimated $g$-band magnitudes from SDSS and PS. To obtain their $g$-band absolute magnitudes we adopted the colour transformation from \citet{Blanton2007},  ${g = B - 0.2354 + 0.3915((g-r) - 0.6102)}$, to convert $B$-band magnitudes to $g$-band magnitudes. The $B$-band magnitudes are computed using the $B$-band galaxy luminosity function of \citet{Marzke1994} with the galaxies' luminosities adopted from \citet{French2020}.
 The galaxy colours are measured using  DESI Legacy Survey imaging in $g$ and $r$ bands. Here, $K$-corrections are not applied for these low redshift galaxies since they are negligible. The measured $g-r$ colour for all the galaxies is presented in Table~\ref{galaxies}.

Using our uniformly computed photometry, we show the g-band absolute magnitude distribution of our galaxies as a function of redshift in Fig.~\ref{fig:sample properties}(c). We find that the vast majority of our galaxies reside above $M_g=-18$ with a few less luminous galaxies. Our sample appears to be mass complete near $M_g=-19$, however this cannot be concluded since this plot is limited to absorbers only and does not account for the {\HI} absorption--mass dependence \citep{bordoloi2018_halomass-HI}. Since this work attempts to eliminate objects with major companions that could have a larger influence on the CGM kinematics, we are less concerned about lower mass companions like LMCs, dwarfs, etc., which can be considered as part of the more massive halo and their kinematics. So having a complete survey to the same depths in each field (i.e., similar central-to-satellite mass ratio limit) is more important than equal mass sensitivity (i.e., $10^9$~M$_{\odot}$) across all fields,  which permits us to examine lower masses at lower redshifts.  We have also verified that our results remain unchanged within the errors reported here when we exclude low luminosity/mass galaxies ($M_g>-19$, 9 galaxies in total). Thus, we include all galaxies in our analysis.

To test the validity of our computed masses, we  compared our values with the stellar masses of the 11 galaxies that overlap with the COS-Halos sample \citep{werk2013}. We found a mean difference of 0.065~dex between the two samples., which provides confidence in our mass estimates. 

We also converted the galaxy stellar masses ($M_{\ast}$) to halo masses ($M_{\rm h}$) using the stellar-to-halo mass relation (SHMR) from \citet{Girelli2020}. We adopted the parameterised SHMR in two redshift bins of $0.0\leq z < 0.2$ and $0.2\leq z < 0.5$. The best-fit parameters with a relative scatter of 0.2~dex from their Table~2 are used for our conversions and uncertainty calculations. The distribution of galaxy masses is shown in Fig.~\ref{fig:sample properties}. The halo masses have a median value of $\log (M_{\rm h}/M_{\odot})=11.5$ and span a full range of $10.5<\log (M_{\rm h}/M_{\odot})<12.7$. The virial radius of all the galaxies in our sample is calculated following the formalism of \citet{Bryan1998}.  The virial radii span a range of $61< R_{\rm vir} < 324$~kpc with a median value of $R_{\rm vir}=131.5$~kpc. The virial radius normalised impact parameters have a range of $0.1\leq D/R_{\rm vir}\leq 5.0$, with a median value of $D/R_{\rm vir}=0.91$. The galaxy masses, $R_{\rm vir}$ and $D/R_{\rm vir}$ can be found in Table~\ref{galaxies}.

\subsection{Galaxy spectroscopy and kinematics}
\label{sec:kine}

\begin{figure}
    \centering
    \includegraphics[width=\columnwidth]{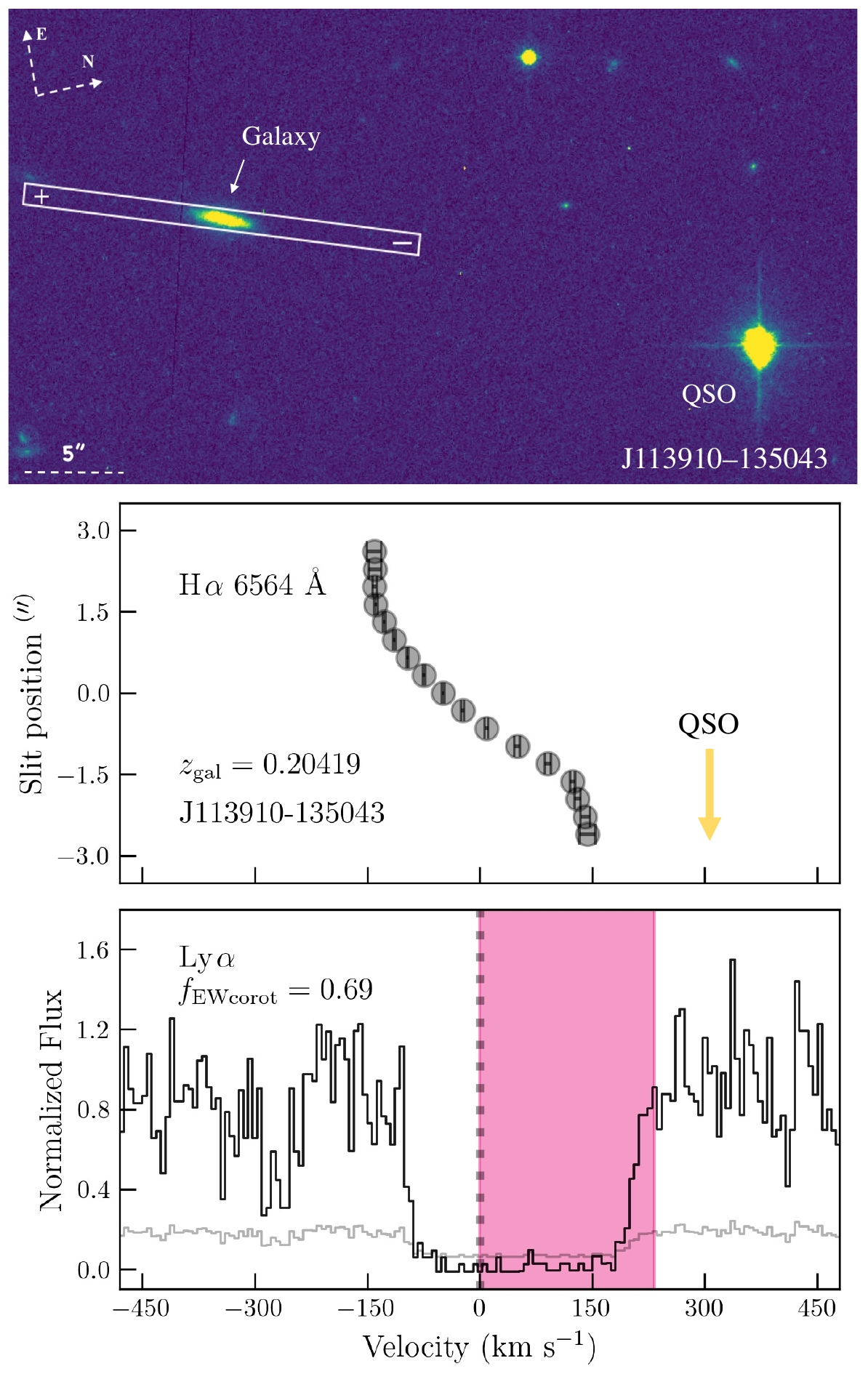} 
    \caption{(Top) {\it HST}/ACS image of the quasar field J$113910-135043$ in the F702W filter. The $1'' \times 20''$ Keck/ESI slit is centred on the galaxy and is aligned with the projected major axis, where the ``+'' and ``--'' signs indicate the positive and negative slit positions. The quasar (QSO) is located $27.6''$ ($D=93.2$~kpc and $D/R_{\rm vir}=0.62$) away from the galaxy on the negative slit position side. (Middle) Extracted {\Ha} rotation curve for the galaxy at $z_{\rm gal}=0.20419$. The galaxy velocities are receding in the direction of the quasar sightline, so any co-rotating {\HI} CGM absorption should also have positive velocities. (Bottom) The {\Lya} absorption profile observed in the background quasar spectrum, where black is the data and grey is the error spectrum. The velocity window of the pink-shaded region is defined to cover the absorption residing to the side of the galaxy systemic velocity (vertical dotted line) corresponding to the galaxy's rotation curve in the direction of the quasar sightline. We integrate over this region to determine the equivalent width co-rotation fraction, which is $f_{\rm EWcorot}=0.69$ for this absorption system.}
 
    \label{fig:methods}
\end{figure}

To compare the CGM kinematics to the kinematics of the galaxies, we require galaxy redshift zeropoints and their rotation curves. The galaxy kinematics for 23 galaxies were obtained from \citet{French2020}.
We further obtained spectra for 39 galaxies using the Keck/ESI over the course of 10 observing nights across 2010, 2014, 2015, and 2016. The wavelength coverage of ESI is 4000 -- 11000~{\AA}, which covers a range of emission lines like the {\OII} doublet, {\Hb}, the {\OIII} doublet, {\Ha}, and {\NII} doublet. The width of the ESI slit was set to $1''$ and it is $20''$ long. The slit position angle was selected to be aligned with the optical major axis of each galaxy to acquire their full range of rotation velocities (see Fig.~\ref{fig:methods}). The echellete spectra obtained over $2014-2016$ were binned on-chip by two in the spatial and spectral directions resulting in a pixel size of  $0\farcs27-0\farcs34$ and spectral resolution of $R\sim4600$ with a sampling rate of 22~{\kms} pixel$^{-1}$ (${\rm FWHM}\sim65$~{\kms}). The spectra obtained in 2010 were binned only spatially on-chip by two. 

The standard echelle package in IRAF was used to combine, to perform flat-field correction, and to extract the ESI spectra. The wavelength solutions were derived using a list of known sky-lines having vacuum wavelengths, where our wavelength solutions have a rms scatter of $\sim0.03$~{\AA} or about 2~{\kms}. The spectra were also heliocentric velocity corrected.

The galaxy rotation curves were extracted following the method described in \citet{Glenn2010a} with a similar approach used by \citet{vogt1996} and \citet{Steidel2002}. In summary, we adopted a three-pixel-wide aperture size and shifted the aperture by one pixel intervals along the spatial direction and extracted a series of spectra along the major axis of each galaxy. We performed Gaussian fits to galaxy emission lines (mainly {\Ha} and {\Hb}), which provided the wavelength centroids used to derive the galaxy systemic redshifts and rotation curves. The galaxy redshifts are listed in Table~\ref{galaxies}.
Fig.~\ref{fig:methods} shows the extracted rotation curve of a galaxy at $z_{\rm gal}=0.20419$ associated with {\HI} absorption in J$113910-135043$, where the {\Ha} emission line was used to extract this rotation curve. In this particular geometry, the quasar sightline is positioned in the negative direction along the slit along the galaxy major axis.

\subsection{Quasar spectroscopy}
Our sample contains 58 quasars, with some quasars probing multiple galaxies and some galaxies having multiple quasar sightlines. The background quasars in each field were observed with the {\it HST}/COS and
Table~\ref{QSOs} provides details of the quasar spectroscopy with the coordinates, redshifts, {\it HST} program IDs, and the COS gratings used. The far-ultraviolet gratings G130M and/or G160M have a moderate resolving power  of $R\sim20,000$,  giving a full width at half maximum of $\sim18$~{\kms} and wavelength coverage of $1410-1780$~{\AA}. We used the STScI CALCOS V2.21 pipeline \citep{Massa_COS} to reduce and flux calibrate all spectral data acquired from the {\it HST} archive. All spectra are heliocentric velocity corrected and in vacuum wavelengths. We co-added multiple integrations with the IDL code {\sc coadd\_x1d}\footnote{\url{http://casa.colorado.edu/~danforth/science/cos/costools.html}} \citep{Danforth10} and binned spectrally by three pixels to enable an increased signal-to-noise ratio. Continuum normalisation was performed by fitting low-order polynomials to the spectra while excluding regions with strong absorption lines.

We implemented the interactive {\sc SYSANAL} code \citep[][]{sysanal_churchill1997,sysanal_churchill&vogt2001} to define the velocity bounds of {\Lya} absorption profiles, compute the optical depth-weighted mean systemic redshifts of absorption ($z_{\rm abs}$), and to compute the rest-frame equivalent widths ($W_r({1215})$). Column densities were adopted from \citet{sameer2024} and are listed in Table~\ref{absorption}. They use a cloud-by-cloud, multi-phase, Bayesian ionisation modelling approach to determine the physical properties of the absorption systems. It has been demonstrated to produce reliable column densities even when a saturated {\Lya} is the only {\HI} line available \citep{sameer2021}. We do note however, that our results do not depend on the accuracy of the {\HI} column densities as we have selected our highest data bins to account for saturation of {\Lya}.  Where column densities were not available in the literature, we used {\sc VPFIT}\footnote{\url{http://www.ast.cam.ac.uk/~rfc/vpfit.html}} \citep{Carswell2014} to measure the {\HI} column densities by fitting Voigt profile models to the absorption lines. We used the appropriate line spread function\footnote{\url{https://www.stsci.edu/hst/instrumentation/cos/performance/spectral-resolution}} at the corresponding lifetime position when fitting the data. The {\HI} column densities for all absorption systems are listed in Table~\ref{absorption}.

Fig.~\ref{fig:sample properties} shows the distribution of rest-frame equivalent widths and column densities as a function of the impact parameter. Both show a strong anti-correlation between the absorption strength and $D$. While high column density systems tend to exist only within the inner halos of galaxies, lower column density systems tend to reside at low and high impact parameters.

\subsection{HI co-rotation fractions}
\label{corot-fraction-criteria1}

We developed a new method for measuring the co-rotation fraction of the {\HI} halo.
For each {\Lya} absorption system,  we compute the fraction of the total equivalent width that is consistent with our co-rotation model. The only dependent choice required is the velocity window we consider when determining whether the gas is consistent with co-rotation. We choose a velocity window that includes all gas from the systemic velocity onward in the direction of galaxy rotation towards the quasar sightline, defined as $f_{\rm EWcorot}$. A value of 1 indicates all of the gas is consistent with a co-rotation scenario, while 0 suggests none of the gas is  consistent with a co-rotation scenario. Errors on the co-rotation fraction were calculated by bootstrapping the errors associated with galaxy redshift and the absorption profile and range from $0.001-0.008$.

Fig.~\ref{fig:methods} (middle) shows the galaxy rotation curve. In this galaxy-quasar pair, the quasar resides on the negative side of the slit position, where the galaxy's rotation is redshifted with respect to its systemic velocity. For our velocity window criterion, we include all the absorption between the galaxy systemic velocity and the most positive velocity of the absorption boundary defined by SYSANAL as highlighted in pink in Fig.~\ref{fig:methods} (bottom). In this case, 69\% of the absorption equivalent width is consistent with a co-rotation model. This value is comparable to other works where they state that the bulk of the absorption is consistent with co-rotation models \citep[e.g.,][]{Ho17}.

Our new method still allows for a comparison to other works even though different variations of kinematic methods are used \citep[e.g.,][]{Steidel2002,Glenn2010a, Ho17, Glenn2019, French2020} since they discuss co-rotation in a binary form and only when the majority/bulk of the gas is consistent with the model is it co-rotating. Here, we can state that the bulk of the absorption is co-rotating when $f_{\rm EWcorot}\geq 0.5$. Our result now provides a quantification of the amount of gas that is consistent with a co-rotation model. However, we do note that the $f_{\rm EWcorot}$ should be considered an upper limit, and it is plausible that the true co-rotation fraction could be lower since there is the possibility of selecting gas at higher velocities than the galaxy maximum rotation velocity.

\section{Dependence of \boldmath{\HI} co-rotation with galaxy properties}
\label{sec:results}
We investigate the kinematic connection between galaxies and their surrounding {\HI} halos to test the scenarios of gas co-rotation and/or accretion through the CGM. For this purpose, we used the galaxies' rotation curves and velocities of the {\Lya} absorption along quasar sightlines. We remind the reader that although a preference towards edge-on galaxies is ideal for examining outflows and inflows, we determined that our results remain unchanged within the errors reported here when we exclude galaxies with $i<30$~deg. Thus, for the remainder of the paper, we include all galaxies in our analysis. In the following sections, we explore the {\Lya} equivalent width co-rotation fractions, $f_{\rm EWcorot}$,  
for a range of properties such as absorption strength, impact parameter, galaxy orientation, and stellar mass.

\subsection{\boldmath$f_{\rm EWcorot}$ and \boldmath{\HI} column density}
\label{corot-NHI}
Simulations show that the CGM has a vast range of {\HI} column densities and that different column density regimes may probe different components of the CGM \citep{Fumagalli2011,suresh2019}. For example, simulations have shown that Lyman Limit Systems (LLSs) may be the best probe of CGM gas flows, including accretion and outflows \citep{vandeVoort2012,FaucherGiguere&keres2011,FaucherGiguere2015,Hafen2017}.  Here we explore how the gas co-rotation fraction of {\Lya} behaves as a function of {\HI} column density. 

\begin{figure}
    \centering
    \includegraphics[width=\columnwidth]{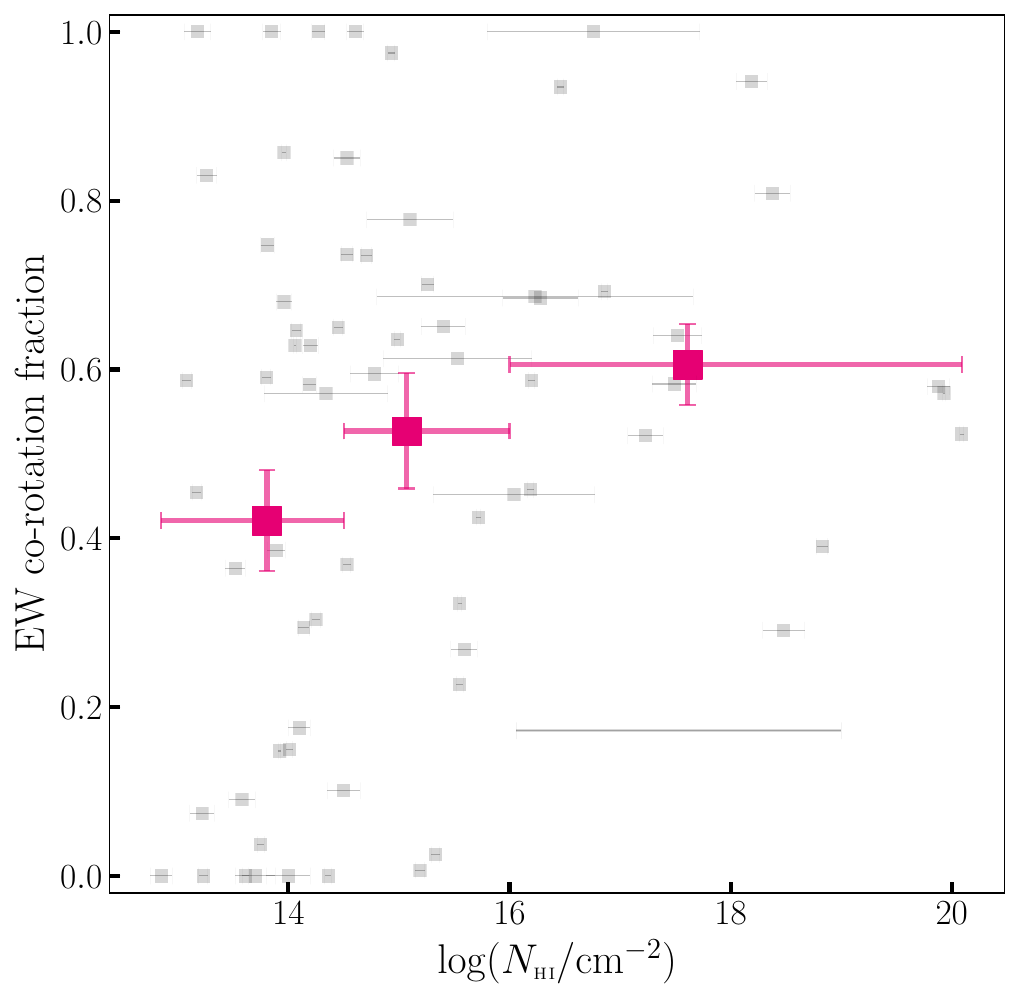}\hfill 
    \caption{Equivalent width co-rotation fraction ($f_{\rm EWcorot}$) as a function of {\HI} column density. The
grey data points are individual galaxies and the grey bars represent the column density errors. The horizontal grey bar spanning $16\leq{\colden}\leq19$ at $f_{\rm EWcorot}\sim0.2$ represents an absorption system where the column density is poorly constrained due to saturation. The pink squares are the averaged $f_{\rm EWcorot}$ in bins of column density, where the error bars represent the column density ranges of each bin and the $1\sigma$ bootstrapped errors on $f_{\rm EWcorot}$. The fraction of {\HI} absorption that is consistent with co-rotation increases with increasing the column density.}
    \label{fig:corot-NHI}
\end{figure}

Figure~\ref{fig:corot-NHI} presents $f_{\rm EWcorot}$ as a function of {\HI} column density.  The grey data points are individual absorption system column densities. We find that the low column density systems span the full range of $f_{\rm EWcorot}$ whereas higher column density systems, particularly those above ${\colden}\sim16$ tend toward higher $f_{\rm EWcorot}$. To better investigate this trend, we divided the data into three bins of column density: ${\colden}<14.5$, $14.5\leq{\colden}<16.2$, and ${\colden}\geq16.2$. The large pink squares represent the mean $f_{\rm EWcorot}$ in each column density bin where the vertical error bars are calculated using 10,000 bootstrapped realisations of the data and their errors to measure the mean and its 1$\sigma$ error. We find that the $f_{\rm EWcorot}$ is correlated with ${\colden}$, where the co-rotation fraction increases from 0.42$\pm$0.06 for the lowest column density bin to 0.59$\pm$0.05 in the highest column density bin. Therefore stronger absorbers are more likely to have kinematics that are consistent with having gas with line of sight velocities aligned with the rotation curve of the galaxy, increasing from 40\% to 60\% co-rotation. 

 \begin{figure*}
    \centering
    \includegraphics[width=\columnwidth]{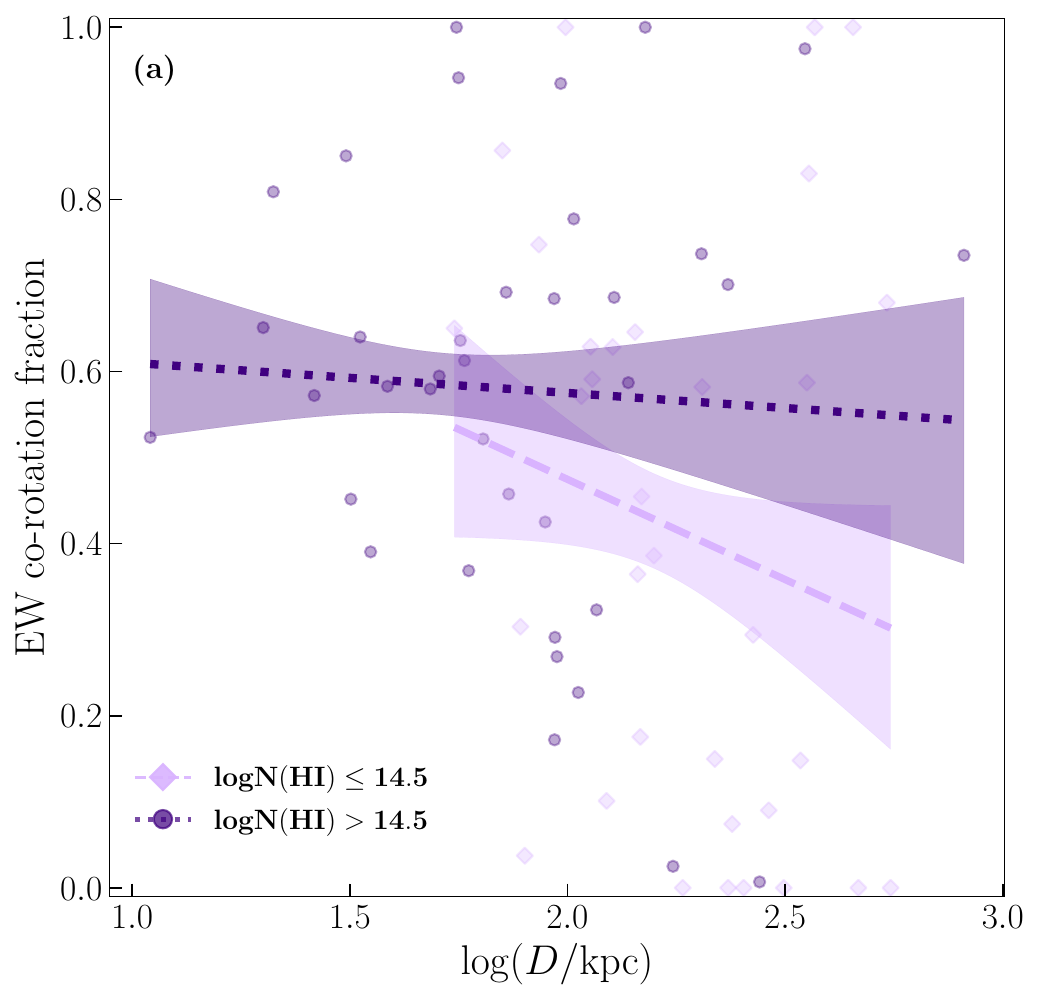}\hfill 
    \includegraphics[width=\columnwidth]{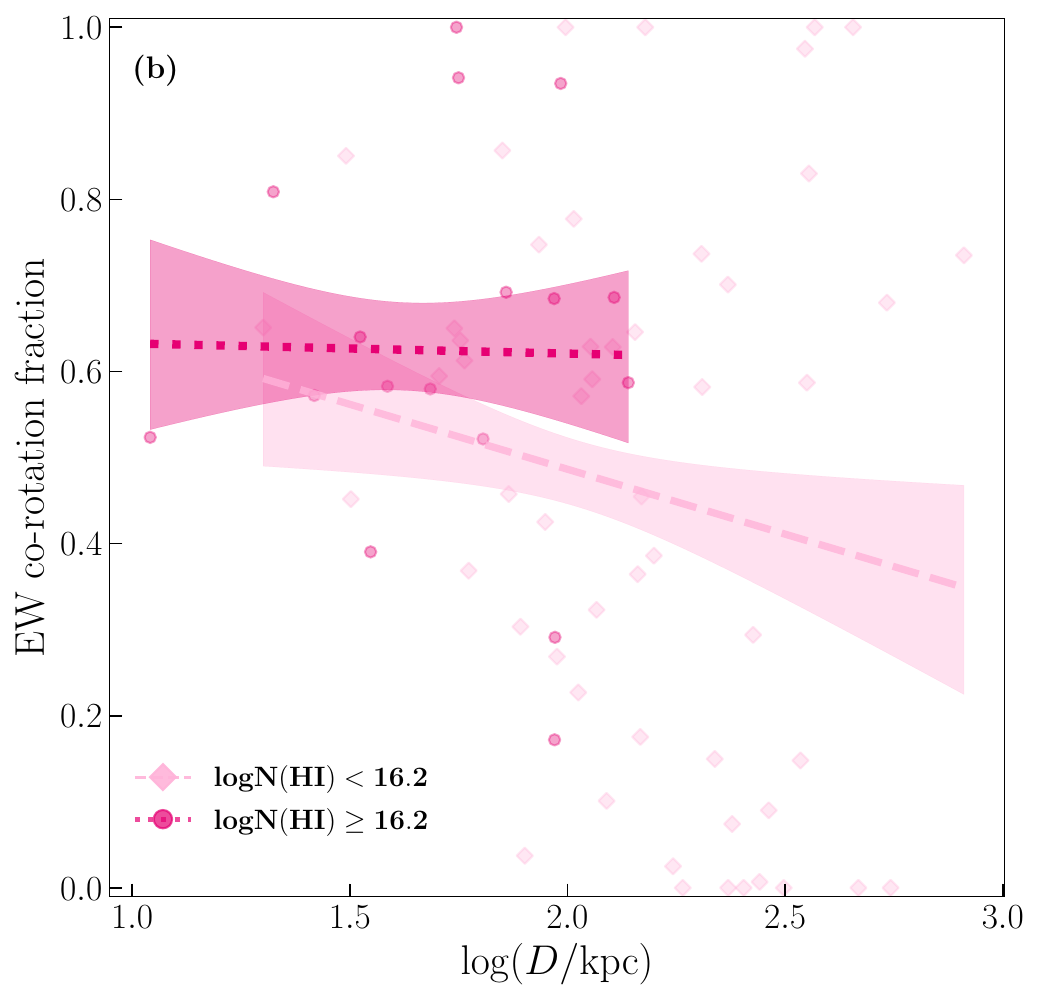}\hfill 
    \caption{Equivalent width co-rotation fraction ($f_{\rm EWcorot}$)  as a function of impact parameter ($D$). (a) The sample is split into low ($\log (N({\HI})/{\rm cm}^{-2})\leq14.5$) and high ($\log (N({\HI})/{\rm cm}^{-2})>14.5$) column density. The slope of the dark purple fit (high column density subsample) is consistent with the light purple fit (low column density systems) within the $1\sigma$ bootstrap errors. (b) The sample is divided into low ($\log (N({\HI})/{\rm cm}^{-2})<16.2$) and high ($\log (N({\HI})/{\rm cm}^{-2})\geq16.2$) column density absorbers. While {\fewcorot} is consistent with having a flat distribution with $D$ for all subsamples, the lower column density subsamples tend towards a decreasing {\fewcorot} as the impact parameter increases.}
    \label{fig:corot-D-NHIcuts}
\end{figure*}

\subsection{\boldmath$f_{\rm EWcorot}$ vs \boldmath$D$ and \boldmath$D/R_{\bf vir}$}
Quasar absorption line studies have shown that {\HI} column densities decrease with increasing impact parameter \citep[e.g.,][]{Tumlinson13,Borthakur15, Glenn2021}. Combined with our previous results showing the co-rotation fraction may also be dependent on column density, we explore how it behaves as a function of impact parameter as well as column density. To do this, we bifurcated our sample into two sub-samples with splits at $\log (N({\HI})/{\rm cm}^{-2})=14.5$ and $\log (N({\HI})/{\rm cm}^{-2})=16.2$. A column density of 14.5 appears to be where the transition occurs between the CGM and IGM \citep[e.g.,][]{Rudie12,Wakker2015,Bouma2021} and so significant differences in the co-rotation fraction above and below this value can indicate whether the galaxy influences its surroundings beyond $R_{\rm vir}$. The higher column density cut at 16.2 was selected since this is the lower limit for partial Lyman limit systems (pLLs) following the classification by \citet{lehner18}, which is where a bimodality in CGM metallicity is the most apparent \citep{Lehner2013, Wotta16} and where simulations suggest that signatures of gas flows may become dominant.

\begin{figure*}
    \centering
    \includegraphics[width=\columnwidth]{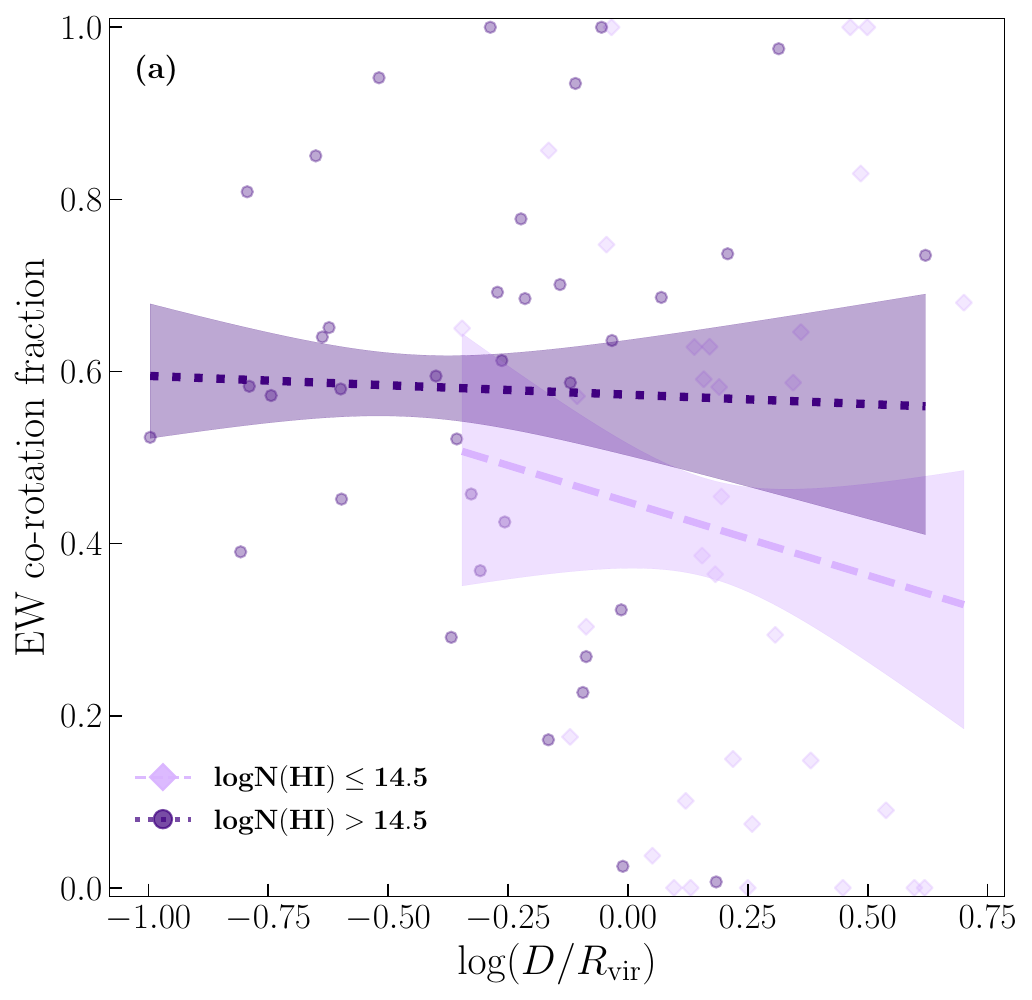}\hfill
    \includegraphics[width=\columnwidth]{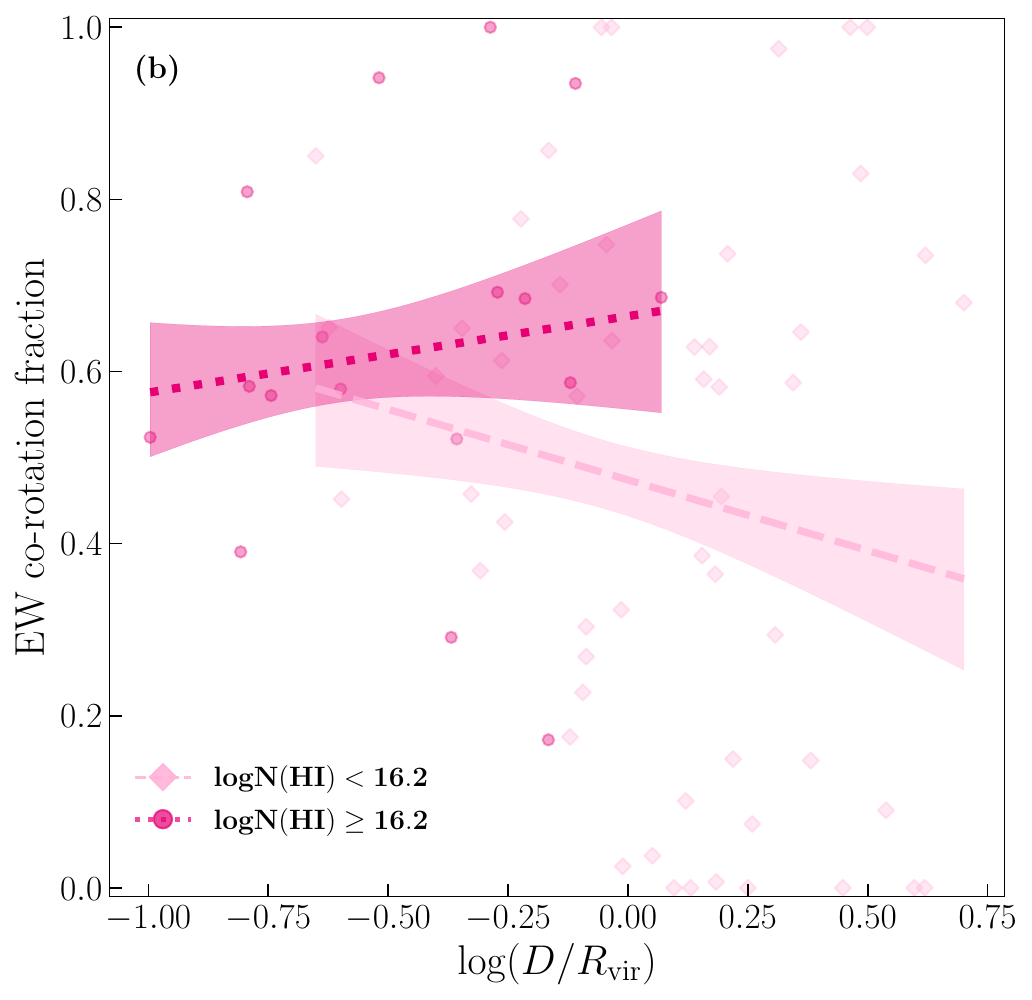}\hfill
    \caption{Same as Fig.~\ref{fig:corot-D-NHIcuts} except the equivalent width co-rotation fraction ($f_{\rm EWcorot}$) is a function of $D/R_{\rm vir}$. (a) For high column density systems (dotted dark purple), the $f_{\rm EWcorot}$ vs $\log(D/R_{\rm vir})$ can be explained with an almost flat distribution. However, the low column density systems (dashed light purple) are decreasing with increasing $D/R_{\rm vir}$. (b) For high column density systems (dotted dark pink), the $f_{\rm EWcorot}$ vs $\log(D/R_{\rm vir})$ can be explained with a slightly increasing distribution, although the curve is still consistent with being flat. The $f_{\rm EWcorot}$ of low column density systems (dashed light pink) decreases with increasing $D/R_{\rm vir}$. Compared to Fig.~\ref{fig:corot-D-NHIcuts}, normalising by $R_{\rm vir}$ affects the slope, where the co-rotation fraction of low column density subsamples decreases with distance more quickly and significantly. This suggests that the mass of the galaxy plays a role in determining whether the gas is co-rotating with the galaxy at a given location.}
        \label{fig:corot-DRvir-NHIcuts}
\end{figure*}

Figure~\ref{fig:corot-D-NHIcuts}(a) shows the first-order polynomial fits to $f_{\rm EWcorot}$ as a function of impact parameter for absorption systems bifurcated by $\log (N({\HI})/{\rm cm}^{-2})=14.5$. The dark and light purple fits present the low (dashed line) and high (dotted line) column density system subsamples, respectively. 
The $1\sigma$ error on the fits is measured by bootstrapping the fit and calculating the average and standard deviation of 10,000 realisations of the data and their errors. 
We find that the $f_{\rm EWcorot}$ for lower column density systems, which reside at larger impact parameters and are more IGM-like, appear to be decreasing but the error bars in the slope are consistent with a flat distribution ($f_{\rm EWcorot} =(-0.17\pm0.26)\log(D/{\rm kpc})+(0.80\pm0.58)$). Higher column density systems, which span a large range of impact parameters, also have a distribution that is consistent with being flat ($f_{\rm EWcorot}=(-0.04\pm0.13)\log(D/{\rm kpc})+(0.65\pm0.22)$).

Fig.~\ref{fig:corot-D-NHIcuts}(b) shows the first-order polynomial fits to $f_{\rm EWcorot}$ for absorption systems bifurcated by $\log (N({\HI})/{\rm cm}^{-2})=16.2$. The light and dark pink fits represent low (dashed line) and high (dotted lines) column densities, respectively. The 1~$\sigma$ errors are also measured with the bootstrapping method. We see that the larger column density systems reside within 100~kpc, while the lower column density systems extend to larger impact parameters. We find that the statistical behaviour of low column density ($f_{\rm EWcorot} = (-0.11\pm0.13)\log(D/{\rm kpc})+(0.72\pm0.27)$), and high column density ($f_{\rm EWcorot} = (-0.01\pm0.16)\log(D/{\rm kpc})+(0.64\pm0.27)$) systems are not significantly different, aside from their extent. The trend in low column density systems (light pink dashed line) shows a slightly decreasing $f_{\rm EWcorot}$ with increasing impact parameter, yet it is consistent with a flat distribution. The high column density systems remain roughly constant with impact parameter (dark pink dotted line).

These trends, or lack thereof, should be taken with caution given that our sample covers a wide range of galaxy masses across 2.5~dex (Fig.~\ref{fig:sample properties}). In fact, the CGM seems to be self-similar over a large mass range \citep{Churchil13_selfsimilar,churchill2013_magii3}, where more massive galaxies host CGM gas out to larger distances, but similar absorption strengths are found at similar fractions of the virial radius across the mass range. Therefore, we normalise the impact parameter by the galaxy virial radius and present the computed values of $D/R_{\rm vir}$ in Table~\ref{galaxies}.

We investigate this mass dependence in Fig.~\ref{fig:corot-DRvir-NHIcuts}(a), which  presents the $f_{\rm EWcorot}$ (purple) versus $D/R_{\rm vir}$ for absorption systems split at a column density of $\log (N({\HI})/{\rm cm}^{-2})=14.5$. We find that higher column density systems (dark purple dotted line) have a flat distribution ($f_{\rm EWcorot}=(-0.02\pm0.13)\log(D/R_{\rm vir})+(0.57\pm0.07)$) that extends beyond the virial radius of the galaxies. For the lower column density systems (light purple dashed line), we find a slightly decreasing trend ($f_{\rm EWcorot}=(-0.20\pm0.26)\log(D/R_{\rm vir})+(0.45\pm0.07)$) where the $f_{\rm EWcorot}$ could decrease beyond the virial radius. Overall, both low and high column density systems are roughly consistent with each other over the overlapping $D/R_{\rm vir}$ range. 

In Fig.~\ref{fig:corot-DRvir-NHIcuts}(b) high column density systems with ${\colden}\geq16.2$ are fitted with a first-order polynomial (dotted dark pink line) that has a positive slope ($f_{\rm EWcorot} = (0.10\pm0.16)\log(D/R_{\rm vir})+(0.66\pm0.11)$), yet is still consistent with a flat distribution. We find a slightly decreasing trend for the low column density systems with ${\colden}<16.2$ plotted as a dashed light pink line ($f_{\rm EWcorot} = (-0.16\pm0.13)\log(D/R_{\rm vir})+(0.47\pm0.04)$), showing that the $f_{\rm EWcorot}$ decreases for absorption detected mostly beyond the virial radius of galaxies.  
This indicates that $f_{\rm EWcorot}$ anti-correlates with impact parameter for low column density absorbers and is flat for high column density systems.

The low significance of these (anti-)correlations could be improved by increasing the sample size. Nevertheless, it seems likely that normalising by the virial radius is an important step in understanding how the gas behaves within galaxy haloes. Regardless of the column density cut selected, a large fraction of the gas ($\sim 60\%$) has kinematics consistent with co-rotation within the virial radius. Outside of the virial radius, there is a decline in the $f_{\rm EWcorot}$.
Thus, a key factor that determines how much of the CGM is co-rotating is its location within the halo and not its column density. 
 
\begin{figure*}
    \centering
    \includegraphics[width=\linewidth]{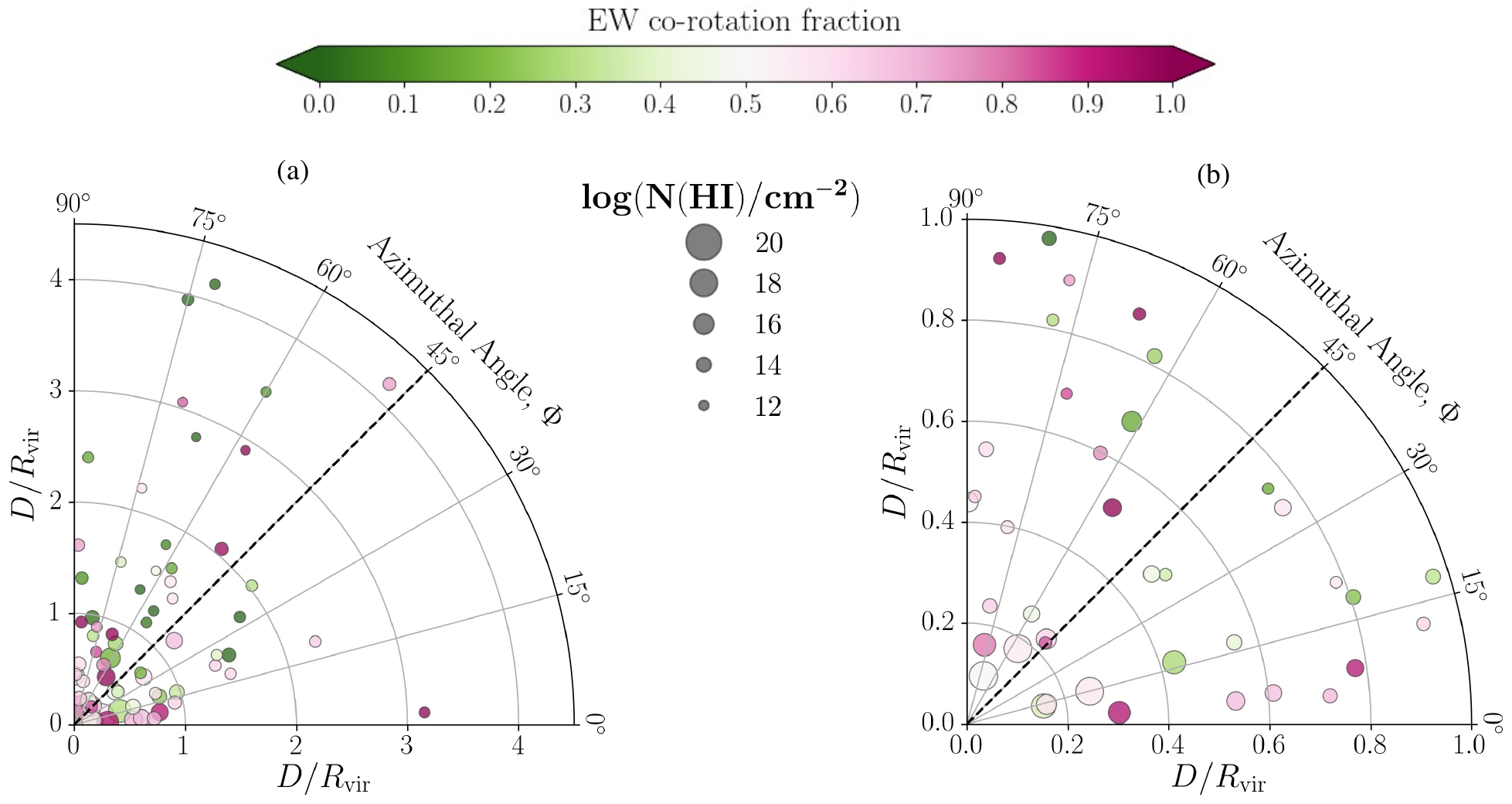}
    \caption{Virial radius normalised impact parameters ($D/R_{\rm vir}$) vs. azimuthal angles ($\Phi$). The host galaxies are located at $D/R_{\rm vir}=0$ with their major axis aligned with $\Phi=0$~deg. Each coloured point represents {\HI} absorption in a background quasar sightline.
  The absorption systems are colour coded by the equivalent width co-rotation fraction ($f_{\rm EWcorot}$) where the darker pink and darker green represent high and low co-rotation fractions, respectively. The point sizes represent the column density of the systems where bigger circles show higher column density and smaller circles show the lower column density absorbers. (a) Distribution of quasar sightlines over the $\Phi$ and $D/R_{\rm vir}$ ranges. We find many systems with high co-rotation fraction~$\geq0.5$ along both the major and minor axes of galaxies. (b) A zoomed-in version of the panel (a). Here we examine systems only within one virial radius of galaxies. We find more systems having kinematics consistent with co-rotation closer to the galaxies along the galaxies' minor axis as well as the major axis.}
    \label{fig:polar}
\end{figure*}

\subsection{\boldmath$f_{\rm EWcorot}$ and galaxy orientation} 

Our current picture of the CGM is one in which cool gas enters the galaxy halo preferentially along the galaxy's major axis and likely accretes onto the galaxy while co-rotating with the disk \citep{stewart2011a, stewart2011b, stewart2013,Nelson2016,stewart2017, suresh2019, peroux2020b}. On the other hand, stellar winds and galaxy feedback will be ejected biconically along the minor axis of the galaxy with higher velocities than the disk \citep{Bouche2012, Schroetter2016, Lan2018, Schroetter19_megaflow3, Bron2022}. Some observations have also shown that the spatial distribution of CGM around galaxies appears to be bimodal \citep{Bordoloi2011, Bouche2012, Glenn2012, Glenn2015decmorpho, Lan2014, Dutta2017,Zabl19_megaflow2}. In order to further examine this picture and the relationships between gas flows with respect to their host galaxies, we explore how the co-rotation fraction relates to galaxy orientation and its behaviour as a function of column density and distance away from the galaxies.

Fig.~\ref{fig:polar} presents $f_{\rm EWcorot}$ as a function of $D/R_{\rm vir}$ and the azimuthal angle, $\Phi$. Panels (a) and (b) are similar; however, while panel (a) shows the full sample, panel (b) plots only {\HI} absorption systems within $R_{\rm vir}$ to focus on gas within the ``halos'' of these galaxies. The host galaxies are located at $D/R_{\rm vir}=0$ with their projected major axis aligned with $\Phi=0$~deg. Each point represents {\HI} absorption in a background quasar sightline and their sizes represent the absorption column densities to emphasise where the higher and lower column densities tend to reside. The anti-correlation between the {\HI} column density and $D/R_{\rm vir}$ is clearly visible (also see Fig.~\ref{fig:sample properties}). The points are also colour-coded based on the measured $f_{\rm EWcorot}$ in each system. The dark pink points in Fig.~\ref{fig:polar} have the highest consistency with co-rotation kinematics, while dark green is least consistent with the host galaxy rotation velocity. 

From the figure, it is clear that the majority (56\%) of our absorption systems reside within $R_{\rm vir}$. On average, the absorption systems beyond $R_{\rm vir}$ tend to have a much lower $f_{\rm EWcorot}$ than within $R_{\rm vir}$, with average {\fewcorot} values of $0.39$ and $0.58$ outside and within $R_{\rm vir}$, respectively. This is consistent with the idea that gas flows may be more organised nearer to galaxies and even that outflows mostly tend not to escape the halos of galaxies \citep{Oppenhimer08,Tumlinson11,Tumlinson13,stocke2013}.  Within $R_{\rm vir}$, absorption systems are consistent with higher $f_{\rm EWcorot}$ and tend to be found within 20~degree of galaxy major and minor axes and cover a large range of {\HI} column densities. Furthermore, these high corotation fractions extend out to $R_{\rm vir}$ along both the major and minor axes. In the remaining part of this subsection, we further 
explore how $f_{\rm EWcorot}$ behaves with azimuthal and inclination angles as a function of column density and $D/R_{\rm vir}$. 

\begin{figure}
    \centering
    \includegraphics[width=\columnwidth]{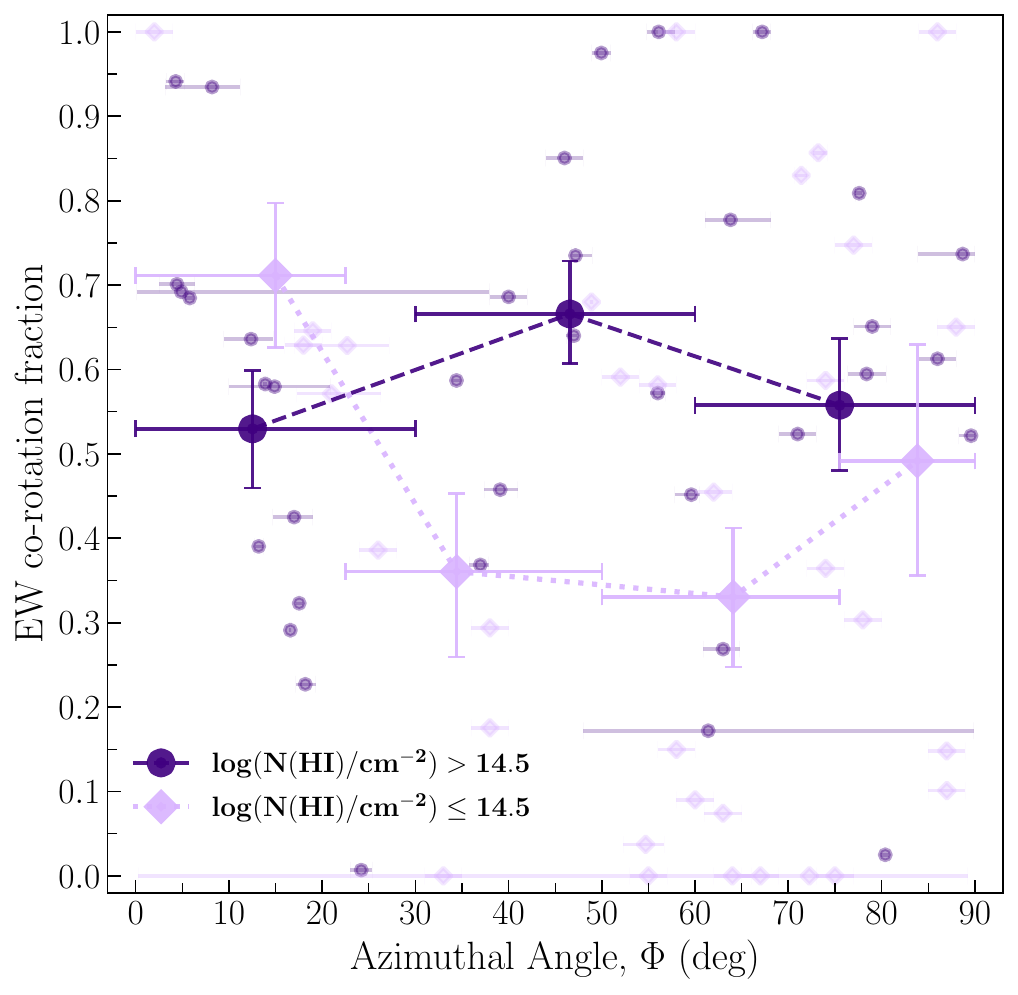}\hfill 
    \caption{Equivalent width co-rotation fraction ($f_{\rm EWcorot}$) as a function of azimuthal angle ($\Phi$) for high (dark purple circles) and low (light purple diamonds) column density systems bifurcated at ${\colden}=14.5$. smaller data points in the background present the individual absorption systems corresponding to each column density cut with the error bars representing the azimuthal angle errors. Absorption systems with ${\colden}>14.5$ in intermediate and high azimuthal angles bins have higher $f_{\rm EWcorot}$, while in lower azimuthal angles, the $f_{\rm EWcorot}$ is larger for absorption systems with ${\colden}\leq14.5$. The $f_{\rm EWcorot}$ is consistent with a flat distribution across all azimuthal angles for the ${\colden}=16.2$ column density cut that is not shown here.}
    \label{fig:corot-AA-NHI}
\end{figure}

\begin{figure}
    \centering
    \includegraphics[width=\columnwidth]{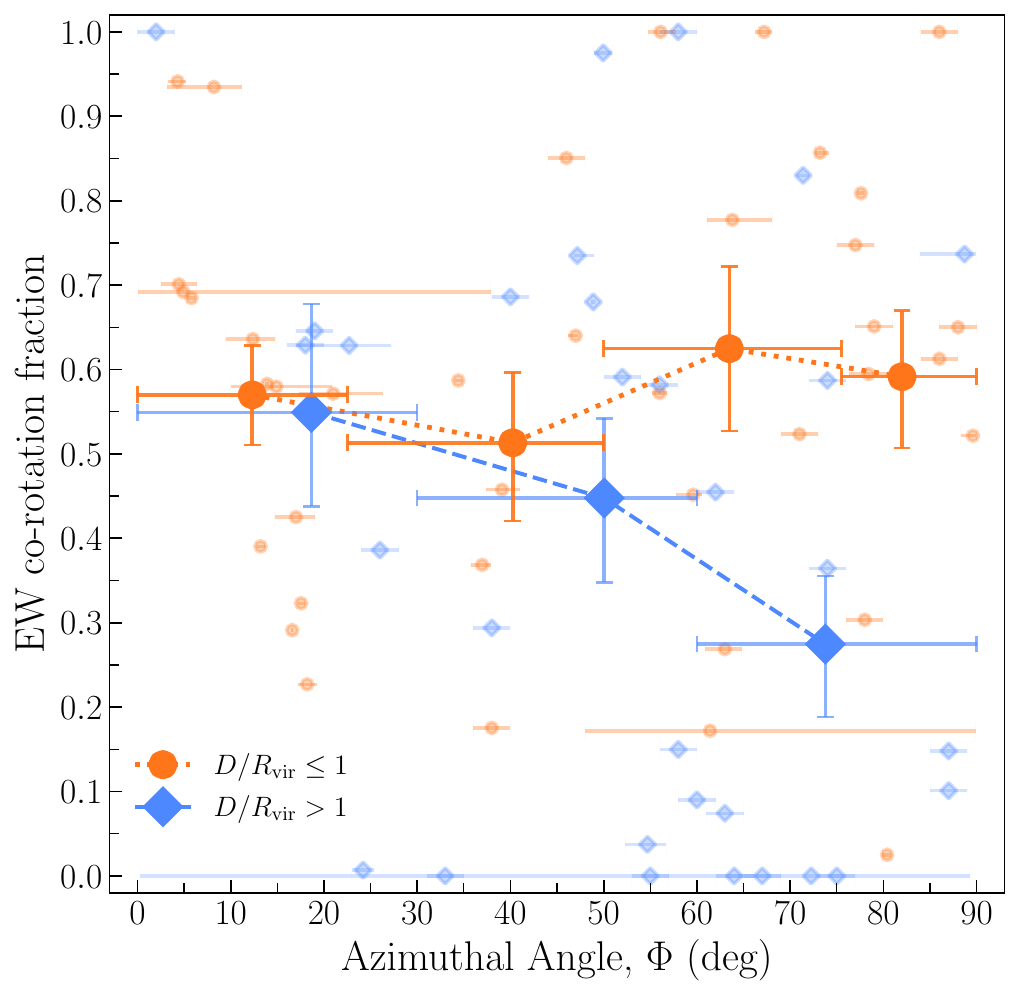}\hfill 
    \caption{Equivalent width co-rotation fraction ($f_{\rm EWcorot}$)  as a function of azimuthal angle ($\Phi$) for $D/R_{\rm vir}>1$ (blue diamonds) and $D/R_{\rm vir}\leq1$ (orange circles). Light orange circles and blue diamonds in the background represent the individual absorbers detected inside and outside the virial radius of host galaxies, respectively. The error bars on coloured data points in the background represent the azimuthal angle errors. While the $f_{\rm EWcorot}$ of  systems within ${R_{\rm vir}}$ increases slightly with increasing $\Phi$, the $f_{\rm EWcorot}$ of {\HI} absorbers with ${D/R_{\rm vir}>1}$ decreases with increasing the azimuthal angle.}
    \label{fig:corot-AA}
\end{figure}

\begin{figure}
        \includegraphics[width=\columnwidth]{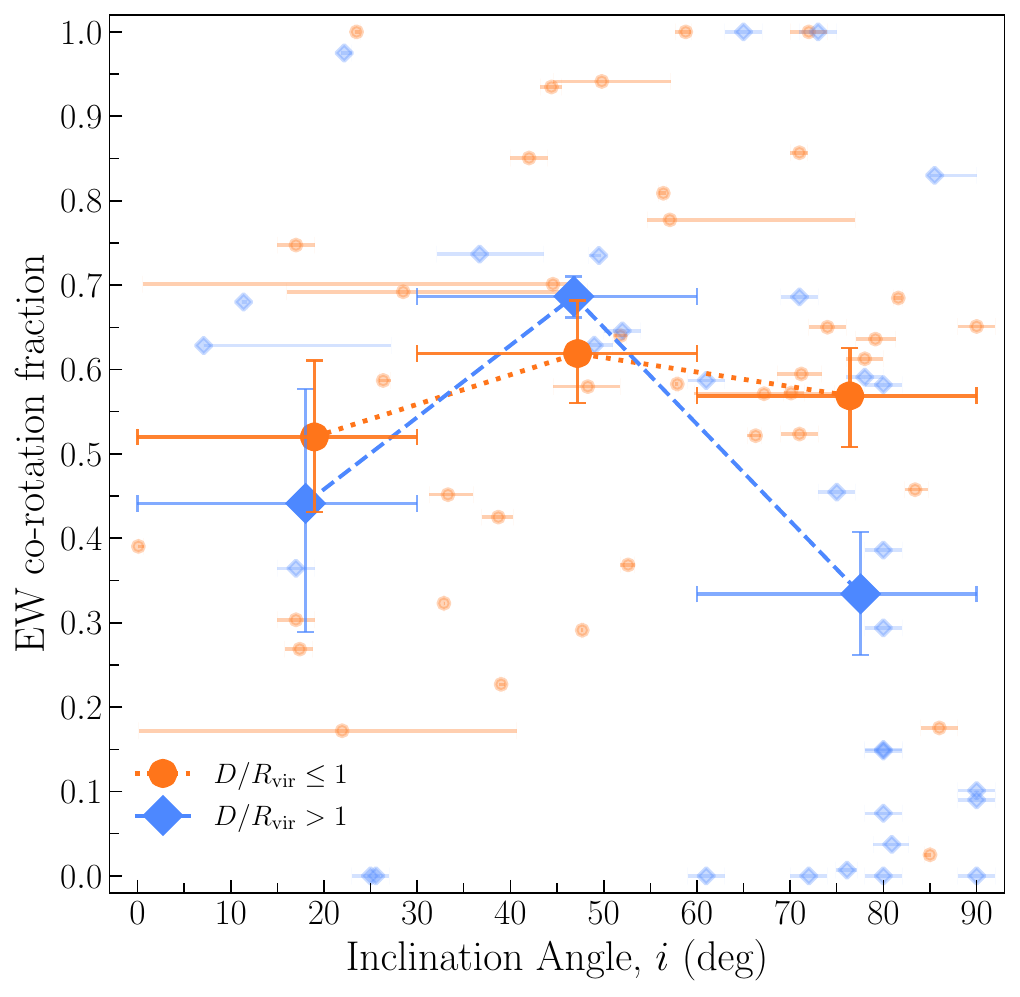}\hfill
            \caption{Equivalent width co-rotation fraction ($f_{\rm EWcorot}$) as a function of inclination angle ($i$) for $D/R_{\rm vir}>1$ (blue diamonds) and $D/R_{\rm vir}\leq1$ (orange circles).The coloured data points in the background are individual galaxies and the error bars represent the inclination angle errors.
            The $f_{\rm EWcorot}$ of {\HI} absorption within ${R_{\rm vir}}$ CGM remains almost constant with inclination angle. The $f_{\rm EWcorot}$ of {\HI} absorbers at ${D/R_{\rm vir}>1}$ decreases for edge-on ($i\geq 60$~degrees) galaxies.} 
    \label{fig:corot-i}   
\end{figure}

We next examine how $f_{\rm EWcorot}$ and azimuthal angle behaves with column density cuts. We apply the same bifurcation in column density of $\log (N({\HI})/{\rm cm}^{-2})=14.5$ and $\log (N({\HI})/{\rm cm}^{-2})=16.2$. Fig.~\ref{fig:corot-AA-NHI} shows the $f_{\rm EWcorot}$ as a function of azimuthal angle with column densities bifurcated at $\log (N({\HI})/{\rm cm}^{-2})=14.5$. The dark purple circles and light purple diamonds present the averaged $f_{\rm EWcorot}$ in bins of azimuthal angles (horizontal bars) for absorbers with ${\log (N({\HI})/{\rm cm}^{-2})>14.5}$ and ${\log (N({\HI})/{\rm cm}^{-2})\leq14.5}$, respectively, with $1\sigma$ bootstrap errors (vertical bars). Here the fainter data points in the background present individual systems where the smaller purple circles show the systems with  higher column densities (${\log (N({\HI})/{\rm cm}^{-2})>14.5}$) and the smaller light purple diamonds present absorbers with lower column densities (${\log (N({\HI})/{\rm cm}^{-2})\leq14.5}$).  The high column density systems only have three bins due to the number of systems and sampling of the azimuthal angles. In general, we find that $f_{\rm EWcorot}$ is higher for high column density systems and lower for low column density systems, except within $\Phi=20-30$~deg of the galaxy major axis where low column density systems have a higher $f_{\rm EWcorot}$. Similarly, we find that the co-rotation fraction ($f_{\rm EWcorot}$) for the $\log (N({\HI})/{\rm cm}^{-2})=16.2$ cut has a higher average value for the high column density systems ($f_{\rm EWcorot}\sim 0.62$) than for the low column density systems ($f_{\rm EWcorot}\sim0.46$). This trend remains flat across all azimuthal angles for the 16.2 column density cut (not shown here). We will discuss the implications of these results in the next section. 

Given that we found $f_{\rm EWcorot}$ is more dependant on $D/R_{\rm vir}$ than $D$ (see Fig.~\ref{fig:corot-DRvir-NHIcuts}), we explore how $f_{\rm EWcorot}$ and azimuthal angle varies with $D/R_{\rm vir}$. In Fig.~\ref{fig:corot-AA} we present the $f_{\rm EWcorot}$ as a function of $\Phi$ where the sample is bifurcated at ${D/R_{\rm vir}=1}$. The orange circles and blue diamonds present the averaged $f_{\rm EWcorot}$ in bins of azimuthal angles (horizontal bars) for absorbers detected at ${D/R_{\rm vir}\leq1}$ and ${D/R_{\rm vir}>1}$, respectively, with $1\sigma$ bootstrap errors (vertical bars). The  smaller data points in the background show individual systems where the blue diamonds are absorption 
detected inside the virial radius and the orange circles are the absorption detected outside the virial.  The data show that along the projected galaxy major axis ($\Phi < 30$~deg), $f_{\rm EWcorot}$ has the same value inside and outside the virial radius with just over half of the gas consistent with co-rotation. This may be expected if gas accretes along filaments, which are co-rotating/accreting from large-scale structure in the IGM down to the galaxy. As the azimuthal angle increases, $f_{\rm EWcorot}$ diverges. We find a high co-rotation fraction for {\HI} gas in the ${R_{\rm vir}}$ CGM (orange) that slightly increases to a peak of $0.6$ along the projected galaxy minor axis ($\Phi>75$~deg). In contrast, {\HI} detected at ${D/R_{\rm vir}>1}$ has a decreasing $f_{\rm EWcorot}$ with increasing $\Phi$, where the value drops to $0.27$ along the projected minor axis. This difference in $f_{\rm EWcorot}$ along the minor axis within and outside the virial radius is a factor of $\sim2$. 

Thus, we find overall that only minor axis absorption (${\Phi\geq60}$~deg) yields significant variations of {\fewcorot} with $D/R_{\rm vir}$, whereas the major axis {\HI} gas appears to have a flat distribution of {\fewcorot} at all radii. If outflows are the dominant source of CGM gas along the minor axis, then outflowing gas co-rotates within the virial radius and either loses angular momentum with increasing height or fails to get to distances beyond the virial radius. 

As our galaxies have a range of inclination angles, we investigate how $f_{\rm EWcorot}$ varies with galaxy inclination. In Fig.~\ref{fig:corot-i} we present $f_{\rm EWcorot}$  as a function of galaxy inclination angle, $i$, where the sample is bifurcated at $D/R_{\rm vir}=1$. We have verified that our sample's distribution of inclination angles is consistent with a random distribution of galaxy inclination angles, and because of this, we have fewer galaxies at low inclination angles. The sample is split into three bins based on the inclination angle of the host galaxies: $i\leq30$~deg, $30<i\leq 60$~deg, and $i>60$~deg. The orange circles show the averaged $f_{\rm EWcorot}$ in each inclination angle bin for systems detected within the $R_{\rm vir}$ of the host galaxies (smaller pale orange data points in the background), while the blue diamonds present the averaged $f_{\rm EWcorot}$ in inclination angle bins for systems detected beyond the virial radius (smaller pale blue diamonds in the background). The vertical bars present the $1\sigma$ bootstrap errors. We find that the $f_{\rm EWcorot}$ of {\HI} absorption within ${R_{\rm vir}}$ remains almost constant across all inclination angles within uncertainties. However, there is a possible trend that beyond $R_{\rm vir}$, $f_{\rm EWcorot}$ drops at high inclination angles when compared to low and intermediate inclination angles. In highly inclined galaxies, $f_{\rm EWcorot}$ within the virial radius is a factor of $\sim2$ higher when compared to {\HI} gas beyond the virial radius. This result could be due to outflows not being able to travel beyond the virial radius and is consistent with what we found in Fig.~\ref{fig:corot-AA} for the azimuthal angle trends. 


\begin{figure*}
    \centering
    \includegraphics[width=\columnwidth]{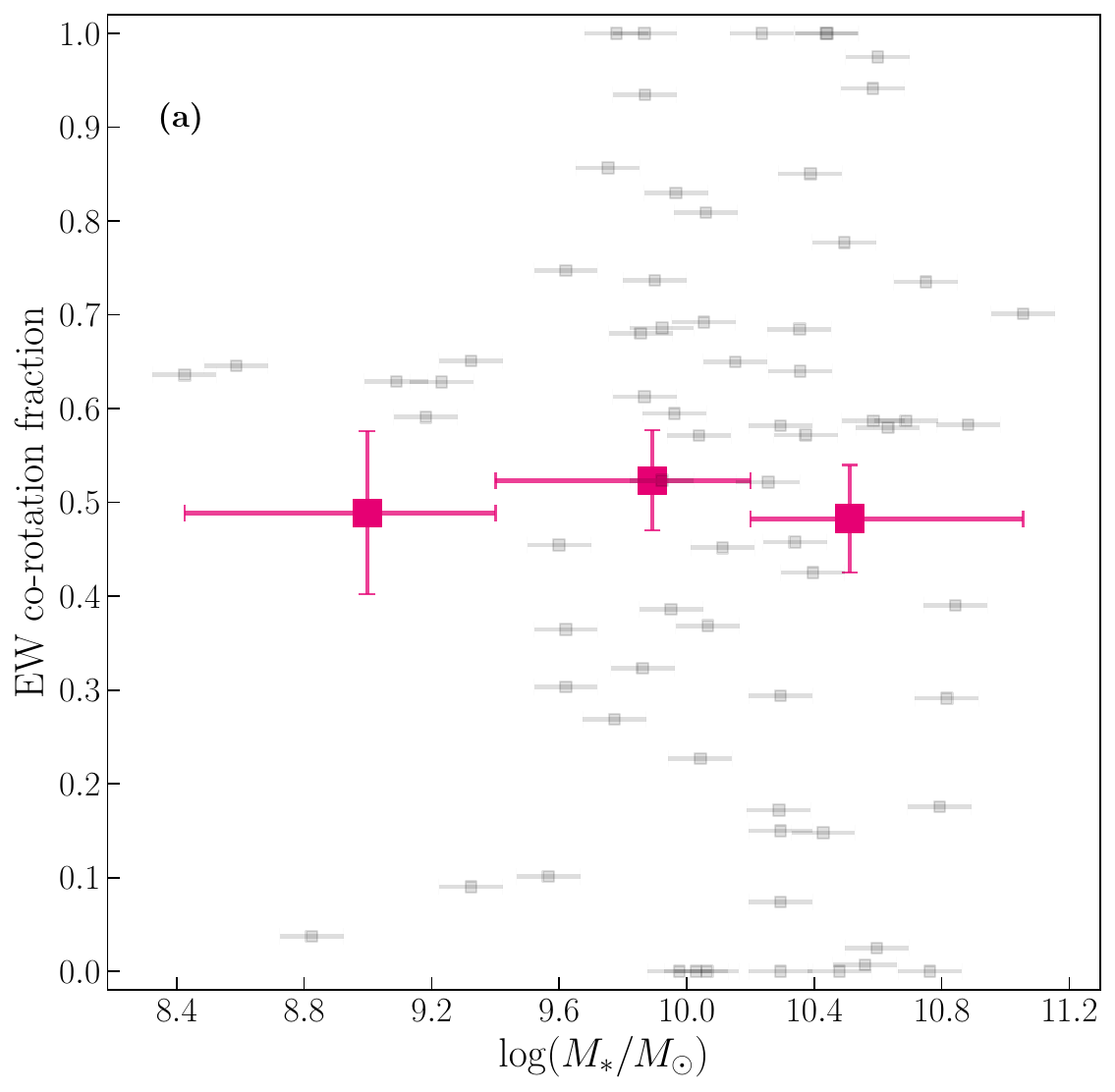}\hfill 
    \includegraphics[width=\columnwidth]{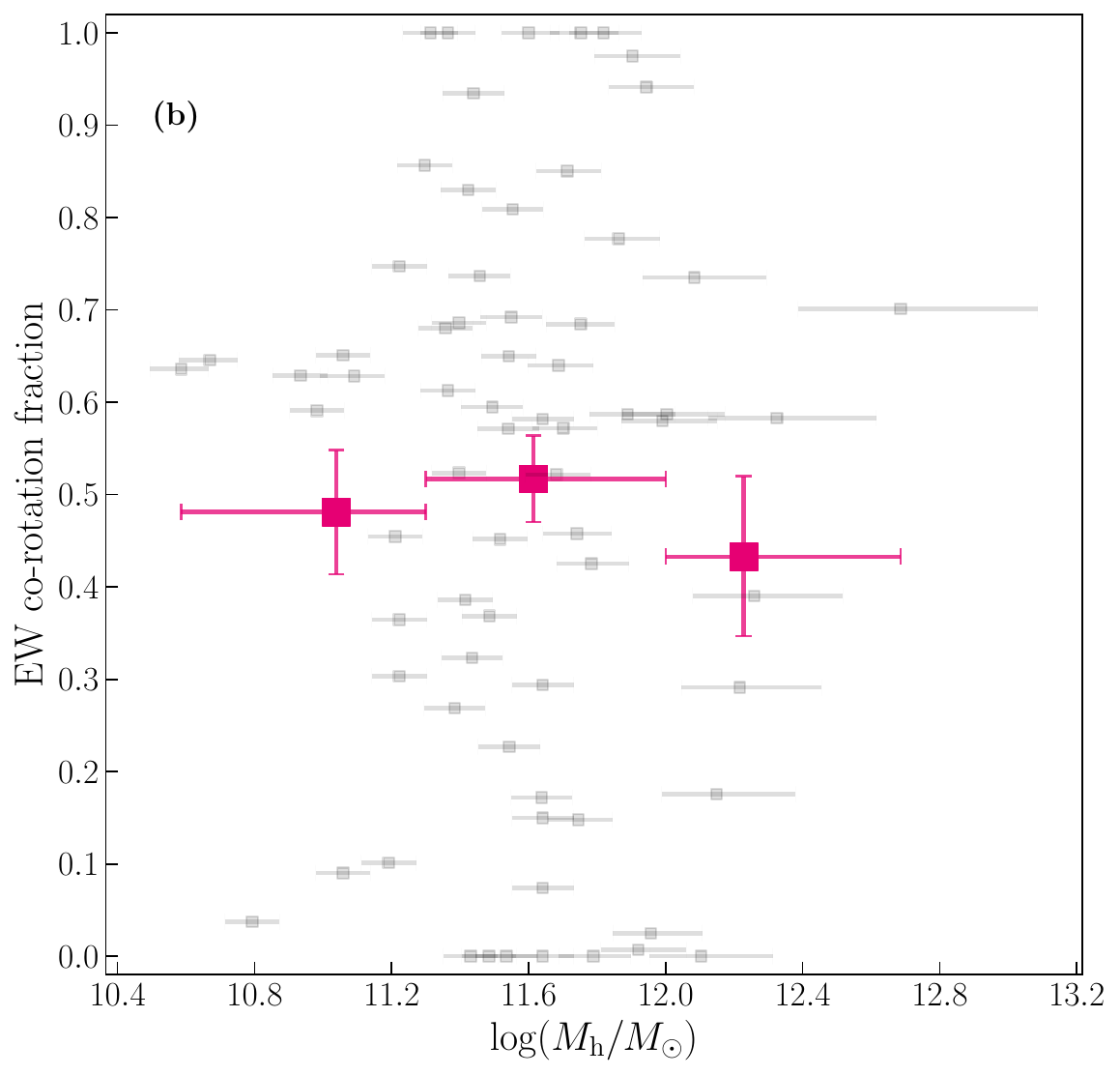}\hfill 
    \caption{(a) {\HI} absorption equivalent width co-rotation fraction ($f_{\rm EWcorot}$) as a function of stellar mass. The grey data points are individual galaxies and the grey
    bars represent the mass errors. The galaxies are split into three stellar mass bins represented by horizontal bars. The pink squares show the averaged $f_{\rm EWcorot}$ in each bin and the $1\sigma$ errors (vertical bars) are measured using a bootstrap method. (b) Same as panel (a) except the $f_{\rm EWcorot}$ is a function of galaxies' halo mass. The equivalent width co-rotation fraction is consistent with a flat distribution across the stellar and halo masses.} 
        \label{corot-Mh}
\end{figure*}

\subsection{$f_{\rm EWcorot}$ and halo mass}
\label{corot-stellarMass}

Some simulations predict that cold-mode accretion halts for halo masses of log($M_{\rm h}/M_{\odot})>12$ \citep{Dekel_Birnboim2006, keres2009,stewart2011a}, which could present as lower {\HI} co-rotation fractions for higher mass galaxies. Therefore, it is important to test how $f_{\rm EWcorot}$ varies with the halo mass. Our sample spans over two decades of halo mass from $10.5<\log (M_{\rm h}/M_{\odot})<12.7$, which allows us to test the simulation prediction. In Fig.~\ref{corot-Mh} (a) and (b) we show the $f_{\rm EWcorot}$ as a function of stellar and halo mass, respectively. The grey data points in the background represent individual galaxies and the grey
bars represent the stellar and halo mass errors, respectively. The pink squares are the averaged $f_{\rm EWcorot}$ in mass bins (horizontal bars) with $1\sigma$ bootstrap errors (vertical bars).  We do not find a significant dependence of $f_{\rm EWcorot}$ on galaxy stellar or halo mass. The data are consistent with being drawn from a flat distribution. Around $\sim$45\% of the {\HI} gas is consistent with a co-rotation model at masses $M_{\rm h}\geq 10^{12} M_{\odot}$, where simulations predict a truncation or halting of cold-mode co-rotating gas accretion. This could imply that the {\HI} is consistent with being coupled to the kinematics of the galaxy at all masses and/or that accretion is present in galaxies of all masses for our sample. It is also possible that the kinematic connection at higher masses is due to the motions of the larger scale environment that those galaxies live in.


\section{Discussion}
\label{sec:discussion}

~{\HI} observations of galactic disks and halos provide new insights into gas flows in the local universe. Feedback, cosmic web filaments, surrounding {\HI} cloud complexes, and minor mergers can all drive the presence and kinematics of gas found in the CGM. It is expected that the connection between the galaxy and CGM kinematics reflects the types of ongoing processes within this diverse gaseous ecosystem. Targeting 70 {\HI} absorption systems assists us in directly probing cool metal-enriched CGM gas over 8 decades of {\HI} column density that is sensitive to a vast range of CGM processes. Using a co-rotating halo assumption to quantify the amount of {\HI} co-rotating gas, we have explored how the kinematics of the CGM relates to the host galaxy properties in an effort to address the origins of this gas. 

\subsection{Interpretation of co-rotation fraction and column density and distance}

The {\HI} column density is an important measure of the CGM as it provides insight into where the gas is located and where it originated from. \citet{Rudie12} determined that ${\colden}<14.5$ most probably traces distant IGM gas while cosmological simulations have shown that ${\colden}>16$ is primarily found in the CGM gas flows within the halos of galaxies \citep[e.g.,][]{Fumagalli2011,suresh2019}. We found that there is a correlation between $f_{\rm EWcorot}$ and N({\HI}) in Fig.~\ref{fig:corot-NHI} where there is an increase in $f_{\rm EWcorot}$ by a factor of $\sim1.5$ from the lowest column density systems to the highest column density systems. This correlation likely implies that there is some dependence on co-rotation and physical processes in a halo, such as outflows from the galaxy, accretion from the IGM, recycled accretion and diffuse components of the CGM. If CGM gas flows exhibit the bulk of the co-rotating gas, then our results are in line with the simulation predictions that higher column density systems are better tracers of gas flows.  However, the {\HI} column density is strongly correlated with the impact parameter and is a hidden additional parameter not accounted for in Fig.~\ref{fig:corot-NHI}.

To examine dependencies with absorption location with respect to the host galaxy, we studied the {\fewcorot} as a function of $D$ and $D/R_{\rm vir}$.
We found no significant difference between the statistical behaviour of low and high column density systems over a large range of impact parameters (see Fig.~\ref{fig:corot-D-NHIcuts}).
However, the scatter in these trends could be significant given the large range of galaxy masses in the sample. We removed the impact of the host galaxies' mass by probing the co-rotation fraction as a function of the virial radius normalised impact parameter (see Fig.~\ref{fig:corot-DRvir-NHIcuts}). We showed that the co-rotation fraction is almost constant within the CGM and drops by a factor of two outside of the virial radii of galaxies. This implies that the {\HI} is most kinematically connected to galaxies within the halo, where most of the physical processes are expected to occur (e.g., outflows, tidal streams, recycling, accretion, etc.).  Thus, both column density and distance from the galaxy play critical roles in where gas is kinematically connected to their galaxies. Nonetheless, there is still 35\% of {\HI} that is consistent with co-rotation out to $3R_{\rm vir}$. This gas could be probing filamentary accretion or larger-scale movements of the local environment.

Our results are consistent with the findings of previous works. \citet{French2020} reported a co-rotation fraction of $59\pm5\%$ for low column density {\Lya} absorbers. Although we use a new method, and our samples overlap at low column densities, the results do not show any significant differences. The authors also reported a decrease in the co-rotation fraction as a function of the impact parameter. However, what we additionally noted here is that both column density and $D/R_{\rm vir}$ are significant factors in the behaviour of $f_{\rm EWcorot}$, with an important transition occurring at $R_{\rm vir}$, where the co-rotation fraction changes from a roughly constant value to rapidly decreasing. 

Our co-rotation fractions are also consistent with those found for {\MgII} absorption systems \citep{Steidel2002,Glenn2010a, Glenn2011june,Ho17,MartinCrystal2019}, which is not so surprising since {\MgII} and {\HI} likely trace similar structures and densities. The main difference between the findings for the two gas tracers is that {\MgII} absorption tends to have the majority of the absorption aligned with the rotation of the galaxy \citep{Steidel2002, Glenn2010a, Ho17}, while {\HI} exhibits a wide range of co-rotation values (see scatter in Fig.~\ref{fig:sample properties}). This could be due to the fact that {\MgII} directly traces higher column density {\HI} that has some metal enrichment where, as we have shown, this higher column density {\HI} has a higher co-rotation fraction. On the other hand, the {\HI} absorption is more spatially widespread than {\MgII}, and traces a larger range of column densities, which could be tracing CGM/IGM or a diffuse component of the CGM, etc., along the same sightline, which has been shown to occur within cosmological simulations \citep{churchill2015,Peeples2019,Marra2021,Marra2022}. 
The lower column density {\HI} can also be traced by {\OVI}, which has lower co-rotation values of $\sim50\%$ \citep{Glenn2019}. We will explore the co-rotation fraction of the metals for our sample in an upcoming paper. Overall, our observations are consistent with those from the literature, which provide a picture of the CGM where we find a stronger kinematic connection to the CGM with higher column densities close to the galaxies and weaker kinematic connection to the CGM with lower column densities further from the galaxies.

\subsection{Interpretation of co-rotation fraction and galaxy orientation}

Our kinematic analysis of {\HI} absorption also supports a non-uniform and likely bi-modal picture of CGM around galaxies. By accounting for the distance away from galaxies, we found that the co-rotation fraction increases with increasing azimuthal angle within $R_{\rm vir}$ and decreases with increasing azimuthal angle beyond $R_{\rm vir}$ (Fig.~\ref{fig:corot-AA}).  This is further supported by a similar trend for galaxy inclination angle, where the co-rotation fraction decreases the most at large inclination angles for gas outside $R_{\rm vir}$ (Fig.~\ref{fig:corot-i}). There appears to be a geometric preference for co-rotating gas around galaxies, especially along the minor axis and within $R_{\rm vir}$. Compared to previous results, \citet{French2020} also reported a sharp decrease in their measure of co-rotation fraction above $i>70$~deg, which is consistent with our results for gas outside $R_{\rm vir}$. They likely only saw a decrease since the majority of their sample is low column density {\HI} gas that resides near-to-outside of the virial radius.

\begin{figure}
    \centering
    \includegraphics[width=\columnwidth]{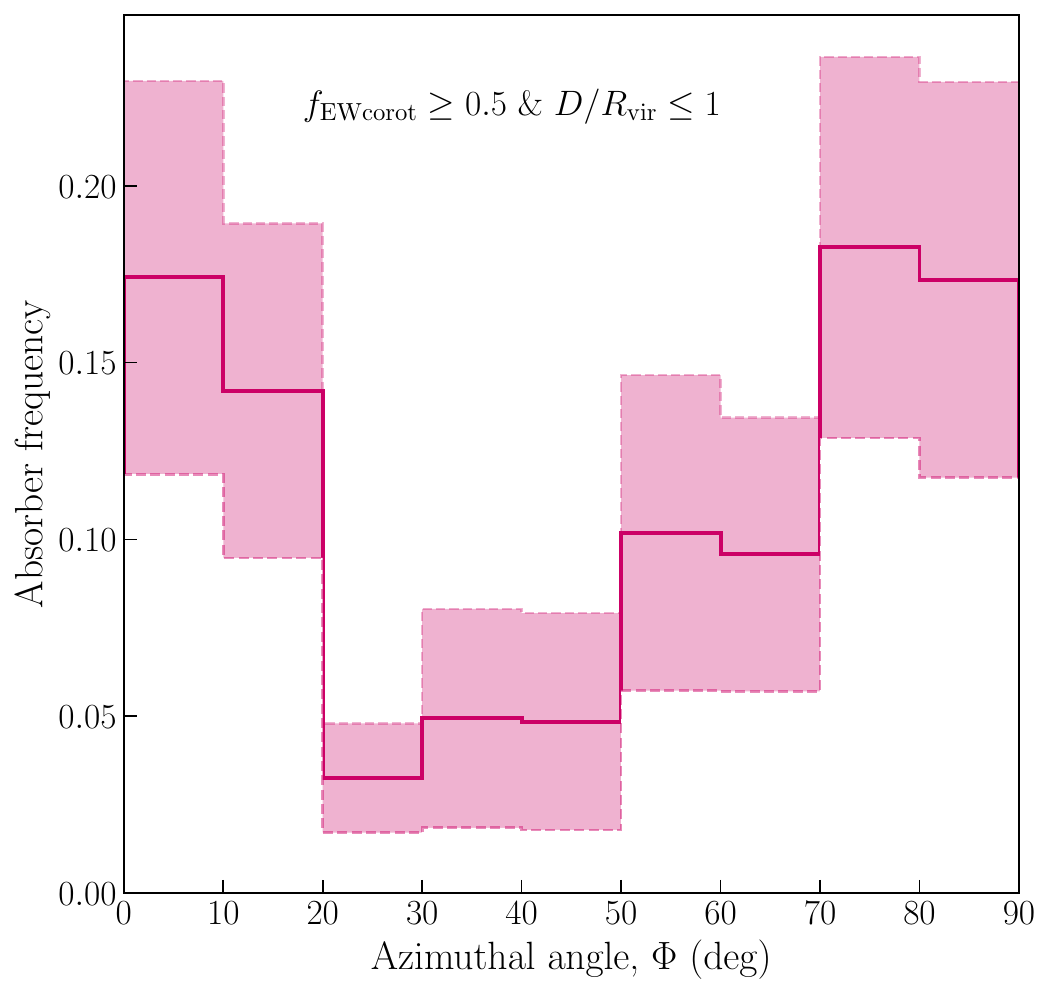}
    \caption{Azimuthal distribution of {\HI} CGM absorption systems within ${{R_{\rm vir}}}$ with high co-rotation fraction (${\fewcorot \geq 0.5 }$). The solid pink line represents the observed frequency in each $\Phi$ bin and the shaded region represents the $1 \sigma$ error measured by bootstrapping the sample. The data suggest a bimodal distribution for CGM absorption with high co-rotation fractions where the histogram peaks along the major and minor axes.}
    \label{fig:bimodal_hist}
\end{figure}

Tying together all of our results and motivations from previous works, we further examine the geometric distribution of co-rotating gas. We computed the frequency of absorption systems as a function of azimuthal angle in Fig.~\ref{fig:bimodal_hist}, focusing on systems dominated by co-rotating gas (e.g., the "bulk" of the gas where $f_{\rm EWcorot}\geq0.5$, which would be similar to other works) within the virial radius of galaxies. Following the methods of \citet{Glenn2012} and \citet{Glenn2015decmorpho}, we model the measured azimuthal angles and their uncertainties for each of the galaxies as asymmetric univariate Gaussian PDFs \citep[see][]{kato2002}. We then compute the mean PDF of all galaxies as a function of $\Phi$. The mean PDF represents the absorption frequency of co-rotating gas at a given $\Phi$. The resulting PDF plotted in Fig.~\ref{fig:bimodal_hist} may be bimodal, where the frequency of highly co-rotating gas within $R_{\rm vir}$ is elevated along the major axis and is highly elevated along the minor axis. This distribution mimics the {\MgII} and {\OVI} covering fraction bi-modalities, which were assumed to be caused by accretion along the major axis and outflows along the minor axis \citep{Glenn2012,Glenn2015decmorpho}. 

The most surprising aspect of our results with azimuthal angle is that we see significant co-rotation along the minor axis, where gas is often assumed to be outflowing from the galaxy. Fig.~\ref{fig:corot-AA} further supports the inference that the minor axis co-rotation is dominated by outflows since we find that $f_{\rm EWcorot}$ diverges with increasing azimuthal angle inside and outside the virial radius. If outflows dominate along the minor axes of galaxies, then they appear to have a higher level of co-rotation within the virial radius, suggesting that outflows travel up to $R_{\rm vir}$, but the drop in co-rotation beyond the virial radius suggests that most outflows do not escape. This is consistent with the relative galaxy--absorption velocities which tend to be below the escape velocity of galaxy halos \citep{stocke2013,Nigel2014}. 

Our co-rotating outflow signatures are in contrast to previous emission line maps of outflowing gas, which tend to show a rapid decrease in the rotation velocities along the minor axis to within a few $10-20$~kpc. Outflows from local starbursting galaxies suggest that signatures of co-rotation diminish beyond 1~kpc from M82 in CO \citep{Leroy2015}. However, rotating gas was found along the outflow axis in recent Enzo (FOGGIE) simulations \citep{Lochhaas2023}. Mapping the gas around a more normal star-forming galaxy in {\MgII}, {\OII}, and {\OIII} emission, \citet{zabl2021} did not find any co-rotation signatures beyond 10~kpc. However, the ionisation mechanism for the {\MgII} and nebular emission is still unknown, so our CGM observations do not necessarily trace the same gas, and the starbursting galaxies have higher star formation rates than our sample, which could result in different outflow kinematic properties. In more comparable work, \citet{MartinCrystal2019} reported no correlation between the sign of the Doppler shift of {\MgII} absorption and the rotation of galaxies along the minor axis. This difference could be due to a difference in {\MgII}/{\HI} gas tracers or most likely, how co-rotation fractions are defined and measured. 
\citet{Glenn2012_dec} reported low metallicity ($\sim -2$dex) multi-phase accretion along the minor axis. They found that the low ionisation ions, like {\MgII} and {\SiII}, were consistent with co-rotation, 
while the higher ionisation features, like {\CIII},
{\CIV}, and {\OVI}, contained both co-rotating gas as well as a fraction of gas that is inconsistent with co-rotation. This again points to a picture where co-rotation likely depends on {\HI} column density.

If gas accretion via filaments is instead driving the co-rotation seen along the minor axis, then we would expect high co-rotation fractions both inside and outside the virial radius since the accretion is originating from the IGM. However, Fig.~\ref{fig:corot-AA} clearly shows this to not be the case. Accretion more likely explains the major axis kinematic and co-rotation fraction trends where we do not see any significant transition in kinematics as a function of $R_{\rm vir}$, which is expected from simulations where co-rotation is seen beyond the virial radius \citep{Danovich2015,stewart2013,stewart2017}. 

From another perspective, the significant drop in co-rotation at larger distances along the minor axis, is interesting in itself. It is intriguing that there is a significant amount of gas that have kinematics opposite to the rotation of galaxies. One possibility is that this gas could arise from ancient debris from previous galaxies interactions, which can produce retrograde orbits or reversed kinematics. For example, N-body simulations of NGC 7252 and NGC 3921 are able to reproduce the reversal kinematics observed in their low {\HI} column density tidal tails \citep{Hibbard_Mihos1995, Hibbard_vanGorkom1996}. Numerical studies of Milky-way also find counter-rotating material around the Galaxy as a result of interactions between disk galaxies. Modelling of interactions implies that merger or fly-by can produce material with retrograde orbits \citep{Pawlowski2011}. This retrograde motion of satellites seems to be quite common in a sample of $z<0.04$ SDSS galaxies, which was shown that 40\% of the galaxies have retrograde motions, which is the same fraction seen in cosmological simulations \citep{Azzaro2006}. Therefore, it is plausible to find low column density gas in retrograde motion at large distances, but it is strange that we see this occurring around the minor axes of our more isolated sample of galaxies. This may only occur if the alignment of the large scale structure is in the direction of the minor axes of these galaxies \citep{Bailin2008}. A larger statistical sample would help to explore this discovery.

Another interesting finding shown in Fig.~\ref{fig:corot-AA-NHI} is that while there is the expected lower $f_{\rm EWcorot}$ for lower column density systems, it is not true along the major axes of galaxies. $f_{\rm EWcorot}$ jumps by nearly a factor of two for lower column density systems, while high column density systems have a lower $f_{\rm EWcorot}$. This could possibly be due to the fragmentation of clouds as they approach the disk or within an accretion stream. One would expect this to occur along the minor axis as well since simulations have shown clouds can fragment as they move through outflows \citep{McCourt2018,sparre2018,Nelson2020}. So it is unclear why we see a transition along the major axis and not the minor axis.  

\subsection{Interpretation of co-rotation fraction and halo mass}

\cite{churchill2013_magii3} studied the relation between {\MgII} equivalent width and host galaxies' virial mass and found no anti-correlation. Despite the prediction of simulations \citep{vandeVoort2011,stewart2011a}, the strength of low-ionisation CGM absorption is not dependent on halo mass and there is no sudden truncation of cold-mode accretion at a mass of $\log (M_{\rm h}/M_{\odot})=12$. It is interesting that our results for {\fewcorot} as a function of $\log (M_{\rm h}/M_{\odot})$ are roughly constant. If we assume that the co-rotating gas is associated with cold-mode accretion, then gas accretion is not dependent on halo mass and cold-mode accretion is still occurring in the most massive galaxy halos. 

It is plausible that the existence of cool gas for high mass galaxies could be arising from their environment, i.e., arising from the galaxies surrounding them.  However, here we find the  majority of the gas within $R_{\rm vir}$ is kinematically consistent with galaxy co-rotation. Given that the vast majority of {\MgII} absorption is found within $R_{\rm vir}$, and that we find most of the gas with $R_{\rm vir}$ is kinematically coupled to the galaxy, then it is less likely that the CGM detected in high mass galaxies arises from larger-scale environments like groups and clusters since they would have larger velocity offsets and could be inconsistent with a co-rotation model. Environmental effects  may be more significant for gas outside $R_{\rm vir}$, where there is  35\% out at $3~R_{\rm vir}$, but this does not represent the bulk of the total absorption seen in the CGM.

\section{Conclusions}
\label{sec:conslusion}
We developed a new method for quantifying the amount of absorption that is consistent with a co-rotation model to examine the kinematic processes within the CGM. In this work, we analysed 70 quasar sightlines with {\HI} absorption detected in their {\it HST}/COS spectra around 62, $z<0.6$, isolated galaxies. Our sample spans a wide range of column densities, $12<\colden<20$, likely associated with both the IGM and CGM. We either measured the rotation curve of galaxies using their Keck/ESI spectra or collected them from literature and connected the spin of galaxies to their {\HI} absorption kinematics. We measured the fraction of {\HI} gas that is aligned with galaxy rotation and examined the {\HI} co-rotation fraction (\fewcorot) as a function of absorption properties such as column density, projected distance from the galaxy ($D$ and $D/R_{\rm vir}$), and its location with respect to galaxies' major axis ($\Phi$), and galaxy properties like inclination angle ($i$) and mass ($\log (M_{\rm h}/M_{\odot})$). Our results include the following:

\begin{enumerate}

    \item The co-rotation fraction of {\HI} absorption is correlated with its column density. {\fewcorot} ranges from $\sim 0.4$ for low column density systems ($\langle \colden \rangle \sim 14$) to $\sim 0.6$ for high column density systems ($\langle \colden \rangle \sim18$). This implies that as the column density increases, the CGM is more coupled to the kinematics of the galaxy and may be more associated with accreting or outflowing gas, however it is important to account for $D/R_{\rm vir}$ in this relation.

\item There is no strong correlation between {\fewcorot} and impact parameter, even when we examine the relation for low and high column density systems. Instead, it is imperative to consider galaxy halo mass and normalise the impact parameter by a galaxy's virial radius when connecting the CGM to its galaxy.  This is because there is a relationship between the {\fewcorot} and $D/R_{\rm vir}$ where there is a flat distribution with ${\fewcorot}\sim 0.6$ within the virial radius. Beyond the viral radius, where the absorption is dominated by lower column density systems, {\fewcorot} decreases within increasing $D/R_{\rm vir}$ and  ${\fewcorot}=0.35$ at the largest distances. These two trends are present when the sample is split by both low (${\colden}=14.5$) and high (${\colden}=16.2$) column densities to separate out the IGM/CGM contributions and the diffuse gas versus gas flow contributions, respectively.

\item Along the major axis ${\fewcorot}\sim0.55$ regardless of distance from the galaxy but this value diverges with increasing azimuthal angle inside and outside the virial radius. Within the virial radius, {\fewcorot} increases to a peak of 0.6 at  $\Phi=90$~deg, while outside of the virial radius, {\fewcorot} decreases to a minimum of 0.27 at $\Phi=90$~deg. If this divergence is caused by outflows, then this could imply that the minor axis gas within $R_{\rm vir}$ is bound and co-rotating with the host galaxy while the gas beyond $R_{\rm vir}$ has less of a co-rotation signature and could be IGM gas. 

\item ~{\fewcorot} shows a similar behaviour with galaxy inclination for face-on galaxies regardless of distance from the galaxy, but the value diverges for edge-on galaxies to ${\fewcorot}=0.6$ for {\HI} within the virial radius and ${\fewcorot}=0.3$ outside the viral radius. This may indicate that the cross-section of outflows decreases outside of the virial radius. 

\item If we examine only {\HI} gas that is dominated by co-rotation (${\fewcorot >0.5}$) and is within the virial radius of galaxies, we find a non-uniform and likely bimodal azimuthal distribution where the gas is preferentially located along the galaxy projected major and minor axes. This result mimics previous covering fraction results with azimuthal angle for both {\MgII} and {\OVI} absorption. Together these findings suggest that gas flows such as accretion and outflows, respectively, are most likely to be found and kinematically connected to host galaxies within $R_{\rm vir}$.

\item There is a significant fraction of co-rotating gas along the minor axis. If this is where outflows are expected, then the outflowing gas maintains rotation out to large fractions of the virial radius. This result is in contrast with previous emission mapping of outflows in {\MgII} and nebular emission for more highly star-forming galaxies, where the gas only co-rotates out to at most $10-20$~kpc. This difference suggests that emission and absorption trace different gas and/or that increased star formation rates reduce the amount of co-rotation in outflows.

\item The {\HI} co-rotation fraction is flat with galaxy stellar and halo mass. This is inconsistent with simulations that predict suppression of {\HI} gas and accretion in massive halos. 

\end{enumerate}

In this work, we examined how the column density and kinematics of {\HI} gas in the CGM relate to galaxy kinematics. We suggest that the different {\HI} column densities probed by {\MgII} and {\OVI} resulted in the different kinematics signatures detected in previous studies. As {\HI} tracks both the low and high ionisation CGM, our results likely explain some of the disparity in previous studies. Thus, {\HI} is likely the best way to study the full range of dynamical processes in the CGM. In the future, we will explore how the metals behave for these systems, especially how the co-rotation fraction changes with different ions. We will also examine how different co-rotation and outflow models affect the co-rotation fraction in an effort to understand how much gas accretion and gas outflow is occurring within halos. 

\section*{Acknowledgements}

We acknowledge the efforts by the Editor who helped improve the quality of the paper. We thank David French for kindly providing the error values used in their paper. H.N, G.G.K, and N.M.N.\ acknowledge the support of the Australian Research Council Centre of Excellence for All Sky Astrophysics in 3 Dimensions (ASTRO 3D), through project number CE170100013. Some of the data presented herein were obtained at the W.~M.~Keck Observatory, which is operated as a scientific partnership among the California Institute of Technology, the University of California and the National Aeronautics and Space Administration. The Observatory was made possible by the generous financial support of the W.~M.~Keck Foundation. Observations were supported by Swinburne Keck programs with 2010A\_W007E, 2010B\_W032E, 2014A\_W178E, 2014B\_W018E, 2015\_W187E, and 2016A\_W056E. The authors wish to recognise and acknowledge the very significant cultural role and reverence that the summit of Maunakea has always had within the indigenous Hawaiian community. We are most fortunate to have the opportunity to conduct observations from this mountain.

\section*{Data Availability}
The data underlying this paper will be shared on reasonable request to the corresponding author.

\bibliographystyle{mnras}
\bibliography{paper2} 


\appendix

\section{Galaxy properties}

Table \ref{galaxies} provides the galaxy details and measurements. In this table, we present the background quasar fields, galaxy ID, coordinates, and redshifts ($z_{\rm gal}$). The galaxy--absorption projected distance ($D$), virial radius normalised impact parameter ($D/R_{\rm vir}$), galaxy inclination angle ($i$), the angle between absorption and galaxy major axis ($\Phi$), galaxy $g-r$ colour, stellar mass ($\log (M_{\ast}/M_{\odot})$), and halo mass ($\log (M_{\rm h}/M_{\odot})$) are also presented in this table.

\begin{landscape}
\begin{table}
\centering
\renewcommand{\arraystretch}{1.2}
\caption{Galaxy properties}
\label{galaxies}

\begin{threeparttable}
\begin{tabular}{l l c c c c c c c c c c c c}
\hline
\hline
\\
  \multicolumn{1}{l}{Quasar} &
  \multicolumn{1}{l}{Galaxy} &
  \multicolumn{1}{c}{RA$_{\rm gal}$} &
  \multicolumn{1}{c}{DEC$_{\rm gal}$} &
  \multicolumn{1}{c}{$z_{\rm gal}$} &
  \multicolumn{1}{c}{$D$ (kpc)} &
  \multicolumn{1}{c}{$D/R_{\rm vir}$} &
  \multicolumn{1}{c}{$i$ (deg)\tnote{a}} &
  \multicolumn{1}{c}{$\Phi$ (deg)\tnote{a}} &
  \multicolumn{1}{c}{$g-r$} &
  \multicolumn{1}{c}{log($M_{\ast}/M_{\odot})$\tnote{b}} &
  \multicolumn{1}{c}{log($M_{\rm h}/M_{\odot})$} &
  \multicolumn{1}{c}{References\tnote{c}}\\
\hline
  MRK335 & NGC7817 & 00:03:58.91 & $+$20:45:08.4 & 0.007702 & 343 & $2.41\substack{+0.08 \\ -0.07}$ & 80$\pm1$ & 87 & 0.87 & 10.43 & $11.75\substack{+0.1 \\ -0.1}$ & 2\\
  J035128$-$142908 & J0351G1 & 03:51:27.87 & $-$14:28:57.9 & 0.356992 & 72.3$\pm$0.4 & $0.53\substack{+0.07 \\ -0.07}$ & $28.5\substack{+19.8\\ -12.5}$ & $4.9\substack{+33 \\ -4.9}$ & 0.29 & 10.05 & $11.55\substack{+0.1 \\ -0.1}$ & 1\\
  PKS0405$-$123 & PKS0405G1 & 04:07:45.63 & $-$12:11:07.1 & 0.361102 & 233.7$\pm$0.4 & $0.72\substack{+0.36 \\ -0.21}$ & $44.6\substack{+2.4 \\ -44.6}$ & $4.4\substack{+ 1.9\\ -1.9}$ & 0.45 & 11.06 & $12.67\substack{+0.4 \\ -0.3}$ & 1\\ 
  J040748$-$121136 & J0407G1 & 04:07:49.67 & $-$12:11:05.5 & 0.495164 & 107.6$\pm$0.4 & $0.78\substack{+0.07 \\ -0.07}$ & $67.2\substack{+ 7.6\\ -7.5}$ & $21.0\substack{+5.3 \\ -3.7}$ & 0.45 & 10.04 & $11.54\substack{+0.1 \\ -0.1}$ & 1\\
  J045608$-$215909 & J0456G1 & 04:56:08.93 & $-$21:59:29.2 & 0.381511 & 103.4$\pm$0.3 & $0.6\substack{+0.1 \\ -0.07}$ & $57.1\substack{+ 19.9\\ -2.4}$ & $63.8\substack{+4.3 \\ -2.7}$ & 0.45 & 10.49 & $11.86\substack{+0.12 \\ -0.1}$ & 1\\
  J045608$-$215909 & J0456G2 & 04:56:09.69 & $-$21:59:03.9 & 0.277938 & 50.7$\pm$0.4 & $0.4\substack{+0.07 \\ -0.07}$ & $71.2\substack{+2.2 \\ -2.6}$ & $78.4\substack{+2.1 \\ -2.0}$ & 0.45 & 9.96 & $11.49\substack{+0.1 \\ -0.1}$ & 1\\
  PG0804$+$761 & UGC04238 & 08:11:36.77 & $+$76:25:17.9 & 0.00515 & 148 & $1.56\substack{+0.06 \\ -0.06}$ & 75$\pm10$ & 62 & 0.45 & 9.59 & $11.21\substack{+0.1 \\ -0.1}$ & 2\\
  J085334$+$434902 & J0853G1 & 08:53:35.16 & $+$43:48:27.3 & 0.09084 & 59.3$\pm$0.1 & $0.49\substack{+0.06 \\ -0.06}$ & $52.6\substack{+0.7 \\ -0.9}$ & $37.0\substack{+0.9 \\ -1.2}$ & 0.53 & 10.07 & $11.49\substack{+0.1 \\ -0.1}$ & 1\\
  J085334$+$434902 & J0853G2 & 08:53:45.24 & $+$43:51:08.2 & 0.163403 & 26.2$\pm$0.1 & $0.18\substack{+0.08 \\ -0.07}$ & $70.1\substack{+ 1.4\\ -0.8}$ & $56.0\substack{+0.8 \\ -0.8}$ & 0.45 & 10.37 & $11.70\substack{+0.1 \\ -0.1}$ & 1\\
  SDSSJ091052$+$333008 & NGC2770 & 09:09:33.71 & $+$33:07:24.7 & 0.006498 & 239 & $1.81\substack{+0.07 \\ -0.07}$ & 80$\pm5$ & 63 & 0.58 & 10.29 & $11.64\substack{+0.1 \\ -0.1}$ & 2\\
  TON1015 & NGC2770 & 09:09:33.71 & $+$33:07:24.7 & 0.006498 & 218 & $1.65\substack{+0.07 \\ -0.07}$ & 80$\pm5$ & 58 & 0.58 & 10.29 & $11.64\substack{+0.1 \\ -0.1}$ & 2\\
  TON1009 & NGC2770 & 09:09:33.71 & $+$33:07:24.7 & 0.006498 & 267 & $2.03\substack{+0.07 \\ -0.07}$ & 80$\pm5$ & 38 & 0.58 & 10.29 & $11.64\substack{+0.1 \\ -0.1}$ & 2\\
  FBQSJ0908$+$3246 & NGC2770 & 09:09:33.71 & $+$33:07:24.7 & 0.006498 & 204 & $1.55\substack{+0.07 \\ -0.07}$ & 80$\pm5$ & 56 & 0.58 & 10.29 & $11.64\substack{+0.1 \\ -0.1}$ & 2\\
  SDSSJ091127$+$325337 & NGC2770 & 09:09:33.71 & $+$33:07:24.7 & 0.006498 & 234 & $1.77\substack{+0.07 \\ -0.07}$ & 80$\pm5$ & 33 & 0.58 & 10.29 & $11.64\substack{+0.1 \\ -0.1}$ & 2\\
  J091440$+$282330 & J0914G1 & 09:14:41.76 & $+$28:23:51.2 & 0.244312 & 105.9$\pm$0.1 & $0.81\substack{+0.07 \\ -0.07}$ & $39.0\substack{+ 0.4\\ -0.2}$ & $18.2\substack{+ 1.1\\ -1.0}$ & 0.17 & 10.04 & $11.54\substack{+0.1 \\ -0.1}$ & 1\\
  J094331$+$053131 & J0943G1 & 09:43:30.72 & $+$05:31:17.5 & 0.353052 & 96.5$\pm$0.3 & $0.78\substack{+0.07 \\ -0.07}$ & $44.4\substack{+ 1.1\\ -1.2}$ & $8.2\substack{+3.0 \\ -5.0}$ & 0.29 & 9.87 & $11.44\substack{+0.1 \\ -0.1}$ & 1\\
  J094331$+$053131 & J0943G2 & 09:43:32.31 & $+$05:31:51.4 & 0.548494 & 150.9$\pm$0.6 & $0.88\substack{+0.09 \\ -0.07}$ & $58.8\substack{+ 0.6\\ -1.1}$ & $67.2\substack{+0.9 \\ -1.0}$ & 0.25 & 10.44 & $11.82\substack{+0.11 \\ -0.1}$ & 1\\
  J095000$+$483129 & J0950G1 & 09:50:01.01 & $+$48:31:02.3 & 0.211866 & 93.6$\pm$0.2 & $0.43\substack{+0.2 \\ -0.12}$ & $47.7\substack{+ 0.1\\ -0.1}$ & $16.6\substack{+0.1 \\ -0.1}$ & 0.45 & 10.82 & $12.22\substack{+0.24 \\ -0.17}$ & 1\\
  PG0953$+$414 & PG0953G1 & 09:57:25.13 & $+$41:20:22.5 & 0.058815 & 541.9$\pm$0.3 & $5.02\substack{+0.06 \\ -0.06}$ & $11.4\substack{+0.4 \\ -0.2}$ & $48.9\substack{+0.2 \\ -0.2}$ & 0.25 & 9.86 & $11.36\substack{+0.1 \\ -0.1}$ & 1\\
  SDSSJ095914$+$320357 & NGC3067 & 09:58:21.08 & $+$32:22:11.6 & 0.004887 & 128 & $1.17\substack{+0.06 \\ -0.06}$ & 71$\pm5$ & 40 & 0.69 & 9.92 & $11.40\substack{+0.1 \\ -0.1}$ & 2\\
  3C232 & NGC3067 & 09:58:21.08 & $+$32:22:11.6 & 0.004887 & 11 & $0.10\substack{+0.06 \\ -0.06}$ & 71$\pm5$ & 71 & 0.69 & 9.92 & $11.40\substack{+0.1 \\ -0.1}$ & 2\\
  PG1001$+$291 & PG1001G1 & 10:04:02.37 & $+$28:55:12.3 & 0.137403 & 56.7 & $0.93\substack{+0.06 \\ -0.07}$ & $79.14\substack{+2.2 \\ -2.1}$ & $12.4\substack{+2.4 \\ -2.9}$ & 0.2 & 8.42 & $10.59\substack{+0.1 \\ -0.1}$ & 1\\
  J100902$+$071343 & J1009G1 & 10:09:02.74 & $+$07:13:37.7 & 0.227855 & 64.0$\pm$0.8 & $0.44\substack{+0.08 \\ -0.07}$ & $66.3\substack{+0.6 \\ -0.9}$ & $89.6\substack{+0.4 \\ -1.3}$ & 0.45 & 10.26 & $11.68\substack{+0.1 \\ -0.1}$ & 1\\
  RX\_J1017.5$+$4702 & NGC3198 & 10:19:54.95 & $+$45:32:58.6 & 0.002202 & 370 & $2.90\substack{+0.07 \\ -0.06}$ & 73$\pm2$ & 58 & 0.56 & 10.24 & $11.60\substack{+0.1 \\ -0.1}$ & 2\\
  J104116$+$061016 & J1041G1 & 10:41:06.32 & $+$06:09:13.5 & 0.442173 & 56.2$\pm$0.3 & 0.30$\substack{+0.07 \\ -0.06}$ & $49.8\substack{+7.4 \\ -5.2}$ & $4.3\substack{+0.9 \\ -1.0}$ & 0.45 & 10.58 & $11.94\substack{+0.14 \\ -0.11}$ & 1\\
  SDSSJ104335$+$115129 & NGC3351 & 10:43:57.70 & $+$11:42:13.7 & 0.002595 & 31 & $0.22\substack{+0.08 \\ -0.07}$ & 42$\pm2$ & 46 & 0.72 & 10.39 & $11.71\substack{+0.1 \\ -0.1}$ & 2\\
  RX\_J1054.2$+$3511 & NGC3432 & 10:52:31.13 & $+$36:37:07.6 & 0.002055 & 290 & $3.45\substack{+0.06 \\ -0.06}$ & 90$\pm4$ & 60 & 0.39 & 9.32 & $11.06\substack{+0.1 \\ -0.1}$ & 2\\
  CSO295 & NGC3432 & 10:52:31.13 & $+$36:37:07.6 & 0.002055 & 20 & $0.24\substack{+0.06 \\ -0.06}$ & 90$\pm2$ & 79 & 0.39 & 9.32 & $11.06\substack{+0.1 \\ -0.1}$ & 2\\
  PG1116$+$215 & PG1116G1 & 11:19:06.70 & $+$21:18:28.8 & 0.138114 & 138.0$\pm0.2$ & $0.76\substack{+0.14 \\ -0.09}$ & $26.4\substack{+0.8 \\ -0.4}$ & $34.4\substack{+0.4 \\ -0.4}$ & 0.45 & 10.69 & $12.00\substack{+0.17 \\ -0.13}$ & 1\\
  PG1116$+$215 & PG1116G2 & 11:19:18.07 & $+$21:15:03.9 & 0.165916 & 814.4$\pm$0.7 & $4.17\substack{+0.17 \\ -0.11}$ & $49.5\substack{+0.2 \\ -1.1}$ & $47.2\substack{+1.8 \\ -0.4}$ & 0.45 & 10.75 & $12.08\substack{+0.21 \\ -0.15}$ & 1\\
  RX\_J1121.2$+$0326 & NGC3633 & 11:20:26.22 & $+$03:35:08.2 & 0.008629 & 184.0 & $1.25\substack{+0.09 \\ -0.07}$ & 72$\pm5$ & 55 & 0.87 & 10.45 & $11.80\substack{+0.11 \\ -0.1}$ & 2\\
  RX\_J1117.6$+$5301 & NGC3631 & 11:21:02.87 & $+$53:10:10.4 & 0.003856 & 78 & $0.82\substack{+0.06 \\ -0.06}$ & 17$\pm5$ & 78 & 0.51 & 9.62 & $11.22\substack{+0.1 \\ -0.1}$ & 2\\
  SDSSJ112448$+$531818 & NGC3631 & 11:21:02.87 & $+$53:10:10.4 & 0.003856 & 86 & $0.90\substack{+0.06 \\ -0.06}$ & 17$\pm5$ & 77 & 0.51 & 9.62 & $11.22\substack{+0.1 \\ -0.1}$ & 2\\
  SDSSJ111443$+$525834 & NGC3631 & 11:21:02.87 & $+$53:10:10.4 & 0.003856 & 145 & $1.52\substack{+0.06 \\ -0.06}$ & 17$\pm5$ & 74 & 0.51 & 9.62 & $11.22\substack{+0.1 \\ -0.1}$ & 2\\
  SDSSJ112114$+$032546 & CGCG039137 & 11:21:26.95 & $+$03:26:41.7 & 0.023076 & 99 & $0.92\substack{+0.06 \\ -0.06}$ & 72$\pm4$ & 86 & 0.61 & 9.87 & $11.36\substack{+0.1 \\ -0.1}$ & 2\\
  SDSSJ112439$+$113117 & NGC3666 & 11:24:26.07 & $+$11:20:32.0 & 0.003546 & 58 & $0.55\substack{+0.06 \\ -0.06}$ & 78$\pm5$ & 86 & 0.61 & 9.87 & $11.36\substack{+0.1 \\ -0.1}$ & 2\\
  SDSSJ112448$+$531818 & UGC06446 & 11:26:40.46 & $+$53:44:48.0 & 0.002151 & 143 & $2.29\substack{+0.06 \\ -0.07}$ & 52$\pm3$ & 19 & 0.31 & 8.59 & $10.67\substack{+0.1 \\ -0.1}$ & 2\\
  J113327$+$032719 & J1133G1 & 11:33:28.27 & $+$03:26:59.6 & 0.154598 & 55.6$\pm$0.1 & $0.52\substack{+0.06 \\ -0.06}$ & $23.5\substack{+0.4 \\ -0.2}$ & $56.1\substack{+1.7 \\ -1.3}$ & 0.2 & 9.78 & $11.31\substack{+0.1 \\ -0.1}$ & 1\\
  \hline

\end{tabular}
\end{threeparttable}
\end{table}
\end{landscape}

\begin{landscape}

\begin{table}

\addtocounter{table}{-1}
\centering
\renewcommand{\arraystretch}{1.2}
\caption{Galaxy Properties continued}

\begin{threeparttable}
\begin{tabular}{l l c c c c c c c c c c c c}
\hline
\hline
\\
  \multicolumn{1}{l}{Quasar} &
  \multicolumn{1}{l}{Galaxy} &
  \multicolumn{1}{c}{RA$_{\rm gal}$} &
  \multicolumn{1}{c}{DEC$_{\rm gal}$} &
  \multicolumn{1}{c}{$z_{\rm gal}$} &
  \multicolumn{1}{c}{$D$ (kpc)} &
  \multicolumn{1}{c}{$D/R_{\rm vir}$} &
  \multicolumn{1}{c}{$i$ (deg)\tnote{a}} &
  \multicolumn{1}{c}{$\Phi$ (deg)\tnote{a}} &
  \multicolumn{1}{c}{$g-r$} &
  \multicolumn{1}{c}{log($M_{\ast}/M_{\odot})$\tnote{b}} &
  \multicolumn{1}{c}{log($M_{\rm h}/M_{\odot})$} & 
  \multicolumn{1}{c}{References\tnote{c}}\\
\\ 
\hline
  J113910$-$135043 & J1139G1 & 11:39:05.90 & $-$13:50:48.1 & 0.219724 & 127.1$\pm$0.1 & $1.38\substack{+0.07 \\ -0.07}$ & $7.1\substack{+20.1 \\ -0.0}$ & $22.7\substack{+4.5 \\ -5.7}$ & 0.45 & 9.23 & $11.09\substack{+0.1 \\ -0.1}$ & 1\\
  J113910$-$135043 & J1139G2 & 11:39:09.52 & $-$13:51:31.8 & 0.212259 & 174.8$\pm$0.1 & $0.98\substack{+0.12 \\ -0.08}$ & $85.0\substack{+0.1 \\ -0.6}$ & $80.4\substack{+0.4 \\ -0.5}$ & 0.78 & 10.6 & $11.96\substack{+0.15 \\ -0.11}$ & 1\\
  J113910$-$135043 & J1139G3 & 11:39:10.01 & $-$13:50:52.3 & 0.319255 & 73.3$\pm$0.4 & $0.47\substack{+0.08 \\ -0.07}$ & $83.4\substack{+1.4 \\ -1.1}$ & $39.1\substack{+1.9 \\ -1.7}$ & 0.45 & 10.34 & $11.74\substack{+0.1 \\ -0.1}$ & 1\\
  J113910$-$135043 & J1139G4 & 11:39:11.53 & $-$13:51:08.6 & 0.204194 & 93.2$\pm$0.3 & $0.61\substack{+0.08 \\ -0.07}$ & $81.6\substack{+0.4 \\ -0.5}$ & $5.8\substack{+0.4 \\ -0.5}$ & 0.66 & 10.35 & $11.75\substack{+0.1 \\ -0.1}$ & 1\\
  PG1216$+$069 & PG1216G1 & 12:19:23.44 & $+$06:38:20.1 & 0.123623 & 93.4 & $0.68\substack{+0.07 \\ -0.07}$ & $22.0\substack{+18.7 \\ -21.8}$ & $61.4\substack{+33 \\ -13.4}$ & 0.41 & 10.29 & $11.64\substack{+0.1 \\ -0.1}$ & 1\\ 
  MRK771 & NGC4529 & 12:32:51.65 & $+$20:11:00.6 & 0.008459 & 158 & $1.43\substack{+0.06 \\ -0.06}$ & 80$\pm8$ & 26 & 0.48 & 9.95 & $11.41\substack{+0.1 \\ -0.1}$ & 2\\
  J123304$-$003134 & J1233G1 & 12:33:03.76 & $-$00:31:59.6 & 0.318757 & 88.9$\pm$0.2 & $0.55\substack{+0.09 \\ -0.07}$ & $38.7\substack{+1.6 \\ -1.8}$ & $17.0\substack{+2.0 \\ -2.3}$ & 0.45 & 10.40 & $11.78\substack{+0.11 \\ -0.1}$ & 1\\
  SDSSJ123604$+$264135 & NGC4565 & 12:36:20.78 & $+$25:59:15.6 & 0.004103 & 147 & $0.76\substack{+0.19 \\ -0.12}$ & 86$\pm7$ & 38 & 0.85 & 10.79 & $12.15\substack{+0.23 \\ -0.16}$ &2\\
  J124154$+$572107 & J1241G1 & 12:41:52.35 & $+$57:20:53.6 & 0.205267 & 21.1$\pm$0.1 & $0.16\substack{+0.07 \\ -0.07}$ & $56.4\substack{+0.3 \\ -0.5}$ & $77.6\substack{+0.3 \\ -0.4}$ & 0.45 & 10.06 & $11.55\substack{+0.1 \\ -0.1}$ & 1\\
  J124154$+$572107 & J1241G2 & 12:41:52.49 & $+$57:20:42.6 & 0.217904 & 94.6$\pm$0.2 & $0.82\substack{+0.07 \\ -0.07}$ & $17.4\substack{+1.4 \\ -1.6}$ & $63.0\substack{+1.8 \\ -2.1}$ & 0.29 & 9.77 & $11.38\substack{+0.1 \\ -0.1}$ & 1\\
  PG1259$+$593 & UGC08146 & 13:02:08.10 & $+$58:42:04.7 & 0.002235 & 114 & $1.44\substack{+0.06 \\ -0.06}$ & 78$\pm3$ & 52 & 0.38 & 9.18 & $10.98\substack{+0.1 \\ -0.1}$ & 2\\
  PG1302$-$102 & NGC4939 & 13:04:14.39 & $-$10:20:22.6 & 0.010317 & 254 & $1.35\substack{+0.17 \\ -0.11}$ & 61$\pm4$ & 64 & 0.57 & 10.76 & $12.10\substack{+0.21 \\ -0.15}$ & 2\\
  J132222$+$464546 & J1322G1 & 13:22:22.51 & $+$46:45:46.0 & 0.214431 & 38.6$\pm$0.2 & $0.16\substack{+0.25 \\ -0.14}$ & $57.9\substack{+0.1 \\ -0.2}$ & $13.9\substack{+0.2 \\ -0.2}$ & 0.69 & 10.88 & $12.32\substack{+0.3 \\ -0.2}$ & 1\\
  J134251$-$005345 & J1342G1 & 13:42:51.76 & $-$00:53:49.3 & 0.227042 & 35.3$\pm$0.2 & $0.16\substack{+0.22 \\ -0.13}$ & $0.1\substack{+0.6\\ -0.1}$ & $13.2\substack{+0.5 \\ -0.4}$ & 0.45 & 10.84 & $12.26\substack{+0.26 \\ -0.18}$ & 1\\
  QSO1500$-$4140 & NGC5786 & 14:58:56.26 & $-$42:00:48.1 & 0.009924 & 453 & $3.15\substack{+0.09 \\ -0.07}$ & 65$\pm5$ & 2 & 0.57 & 10.44 & $11.75\substack{+0.11 \\ -0.1}$ & 2\\
  SDSSJ151237$+$012846 & UGC09760 & 15:12:02.44 & $+$01:41:55.5 & 0.006985 & 123 & $1.32\substack{+0.06 \\ -0.06}$ & 90$\pm4$ & 87 & 0.43 & 9.57 & $11.19\substack{+0.1 \\ -0.1}$ & 2\\
  2E1530$+$1511 & NGC5951 & 15:33:43.06 & $+$15:00:26.2 & 0.005937 & 55 & $0.45\substack{+0.06 \\ -0.06}$ & 74$\pm6$ & 88 & 0.56 & 10.15 & $11.54\substack{+0.1 \\ -0.1}$ & 2\\
  J154743$+$205216 & J1547G1 & 15:47:45.70 & $+$20:49:17.6 & 0.096499 & 79.8$\pm$0.5 & $1.13\substack{+0.06 \\ -0.06}$ & $80.9\substack{+1.8 \\ -2.0}$ & $54.7\substack{+2.0 \\ -2.4}$ & 0.45 & 8.82 & $10.79\substack{+0.1 \\ -0.1}$ & 1\\
  J155504$+$362847 & J1555G1 & 15:55:05.27 & $+$36:28:48.1 & 0.189201 & 33.4$\pm$0.1 & $0.23\substack{+0.08 \\ -0.07}$ & $51.8\substack{+0.7 \\ -0.7}$ & $47.0\substack{+0.3 \\ -0.8}$ & 0.32 & 10.36 & $11.69\substack{+0.1 \\ -0.1}$ & 1\\
  MRK876 & NGC6140 & 16:20:58.16 & $+$65:23:26.0 & 0.003035 & 113 & $1.48\substack{+0.06 \\ -0.06}$ & 49$\pm4$ & 18 & 0.43 & 9.09 & $10.93\substack{+0.1 \\ -0.1}$ & 2\\
  H1821$+$643 & H1821G1 & 18:21:54.53 & $+$64:20:09.0 & 0.225111 & 116.6 & $0.97\substack{+0.07 \\ -0.07}$ & $32.9\substack{+0.04 \\ -0.04}$ & $17.5\substack{+0.4 \\ -0.3}$ & 0.62 & 9.86 & $11.44\substack{+0.1 \\ -0.1}$ & 1\\
  J213135$-$120704 & J2131G1 & 21:31:38.87 & $-$12:06:44.1 & 0.43020 & 48.4$\pm$0.2 & $0.25\substack{+0.13 \\ -0.09}$ & $48.3\substack{+3.5 \\ -3.7}$ & $14.9\substack{+6 \\ -4.9}$ & 0.45 & 10.63 & $12.0\substack{+0.16 \\ -0.12}$ & 1\\
  RBS1768 & ESO343G014 & 21:37:45.18 & $-$38:29:33.2 & 0.030484 & 466 & $3.96\substack{+0.06 \\ -0.06}$ & 90$\pm5$ & 75 & 0.57 & 10.06 & $11.48\substack{+0.1 \\ -0.1}$ & 2\\
  J213745$-$143255 & J2137G1 & 21:37:50.50 & $-$14:30:03.2 & 0.075451 & 70.9$\pm$0.7 & $0.68\substack{+0.06 \\ -0.06}$ & $71.0\substack{+0.9 \\ -1.0}$ & $73.2\substack{+1.0 \\ -0.5}$ & 0.45 & 9.75 & $11.29\substack{+0.1 \\ -0.1}$ & 1\\
  PHL1811 & PHL1811G2 & 21:54:54.66 & $-$09:23:25.39 & 0.325424 & 552.6$\pm$0.8 & $4.16\substack{+0.07 \\ -0.07}$ & 25.6$\substack{+2.7 \\ -4.5}$ & $72.3\substack{+0.17 \\ -0.72}$ & 0.45 & 10.03 & $11.54\substack{+0.1 \\ -0.1}$ & 1\\
  PHL1811 & PHL1811G1 & 21:54:54.93 & $-$09:23:31.1 & 0.176097 & 351.3$\pm$0.3 & $2.06\substack{+0.11 \\ -0.08}$ & $22.17\substack{+0.8 \\ -0.3}$ & $49.9\substack{+1.0 \\ -1.0}$ & 0.45 & 10.59 & $11.90\substack{+0.14 \\ -0.11}$ & 1\\
  PHL1811 & PHL1811G3 & 21:55:05.14 & $-$09:24:25.9 & 0.157933 & 358.8$\pm$0.9 & $3.06\substack{+0.06 \\ -0.06}$ & $85.5\substack{+ 4.5\\ -0.5}$ & $71.4\substack{+0.6 \\ -0.7}$ & 0.43 & 9.97 & $11.42\substack{+0.1 \\ -0.1}$ & 1\\
  MRC2251$-$178 & MCG0358009 & 22:53:40.85 & $-$17:28:44.0 & 0.030071 & 355 & $2.21\substack{+0.11 \\ -0.08}$ & 61$\pm4$ & 74 & 0.63 & 10.58 & $11.89\substack{+0.14 \\ -0.11}$ & 2\\
  J225357$+$160853 & J2253G1 & 22:53:57.80 & $+$16:09:05.5 & 0.153718 & 31.8$\pm$0.2 & $0.25\substack{+0.06 \\ -0.06}$ & $33.3\substack{+2.7 \\ -2.0}$ & $59.6\substack{+0.9 \\ -1.8}$ & 0.45 & 10.11 & $11.52\substack{+0.1 \\ -0.1}$ & 1\\
  J225357$+$160853 & J2253G2 & 22:54:00.37 & $+$16:09:06.4 & 0.352787 & 203.2$\pm$0.5 & $1.61\substack{+0.07 \\ -0.07}$ & $36.7\substack{+6.9 \\ -4.6}$ & $88.7\substack{+1.3 \\ -4.8}$ & 0.08 & 9.90 & $11.46\substack{+0.1 \\ -0.1}$ & 1\\
  J225357$+$160853 & J2253G3 & 22:54:02.32 & $+$16:09:33.4 & 0.390012 & 276.3$\pm$0.2 & $1.53\substack{+0.11 \\ -0.08}$ & $76.1\substack{+1.1 \\ -1.2}$ & $24.2\substack{+1.2 \\ -1.2}$ & 0.45 & 10.56 & $11.92\substack{+0.14 \\ -0.11}$ & 1\\
  RBS2000 & IC5325 & 23:28:43.43 & $-$41:20:00.5 & 0.005043 & 314 & $2.81\substack{+0.06 \\ -0.06}$ & 25$\pm4$ & 67 & 0.57 & 9.98 & $11.43\substack{+0.1 \\ -0.1}$ & 2\\
\hline
\end{tabular}
\begin{tablenotes}
            \item[a] We adopted the inclination angle errors from \cite{French2020}, which were provided by David French (2024, private communication). We note that they used a 3 degree galaxy PA error, which we also adopt here.
            \item[b] We adopt a 0.1 dex error in the stellar masses given the scatter quoted in \citet{Bell2003}. This error is propagated  through to the halo mass errors and $R_{\rm vir}$ errors.
            \item[c] Galaxy kinematics measurements reference: (1) this work, (2) \cite{French2020}.
        \end{tablenotes}
\end{threeparttable}
\end{table}
\end{landscape}

\section{Quasar field observations}

Here we present the quasar information and observations details. Table~\ref{QSOs} includes nine columns and provides the QSO coordinates, redshift, COS observation proposal IDs, the grating(s) used for observation, photometry imager or survey, the filter used for imaging, and the {\it HST} imaging proposal ID. 

\begin{table*}
\centering
\renewcommand{\arraystretch}{1.08}
\caption{QSO observations}

\begin{tabular}{l c c c c c c c c}
\hline
\hline
\\
  \multicolumn{1}{l}{Quasar} &
  \multicolumn{1}{c}{RA (J2000)} &
  \multicolumn{1}{c}{DEC (J2000)} &
  \multicolumn{1}{c}{$z_{\rm qso}$} &
  \multicolumn{1}{c}{COS PID(s)} &
  \multicolumn{1}{c}{COS Gratings} &
  \multicolumn{1}{c}{Imager/Survey} &
  \multicolumn{1}{c}{Filter} &
  \multicolumn{1}{c}{HST PID}\\
\hline
  MRK335 & 00:06:19.5 & $+$20:12:11.0 & 0.026 & 13814 & G130M & ... & ... & ...\\
  J035128$-$142908 & 03:51:28.5 & $-$14:29:08.7 & 0.616 & 13398 & G130M, G160M & HST/WFPC2 & F702W & 5949\\
  PKS0405$-$123 & 04:07:48.4 & $-$12:11:36.7 & 0.573 & 11508 & G130M, G160M & HST/WFPC2 & F702W & 5949\\
  J040748$-$121136 & 04:07:48.4 & $-$12:11:36.7 & 0.572 & 11541 & G130M, G160M & HST/WFPC2 & F702W & 5949\\
  J045608$-$215909 & 04:56:08.9 & $-$21:59:09.4 & 0.533 & 12466,12252,13398 & G160M & HST/WFPC2 & F702W & 5098\\
  PG0804$+$761 & 08:10:58.7 & $+$76:02:43.0 & 0.102 & 11686 & G130M, G160M & ... & ... & ...\\
  J085334$+$434902 & 08:53:34.2 & $+$43:49:02.3 & 0.514 & 13398 & G130M, G160M & HST/WFPC2 & F702W & 5949\\
  FBQSJ0908$+$3246 & 09:08:38.8 & $+$32:46:20.0 & 0.26 & 14240 & G130M & ... & ... & ...\\
  TON1009 & 09:09:06.2 & $+$32:36:30.0 & 0.81 & 12603 & G130M & ... & ... & ...\\
  TON1015 & 09:10:37.0 & $+$33:29:24.0 & 0.354 & 14240 & G130M & ... & ... & ...\\
  SDSSJ091052$+$333008 & 09:10:52.8 & $+$33:30:08.0 & 0.116 & 14240 & G130M & ... & ... & ...\\
  SDSSJ091127$+$325337 & 09:11:27.3 & $+$32:53:37.0 & 0.29 & 14240 & G130M & ... & ... & ...\\
  J091440$+$282330 & 09:14:40.4 & $+$28:23:30.6 & 0.735 & 11598 & G130M, G160M & HST/ACS & F814W & 13024\\
  J094331$+$053131 & 09:43:31.6 & $+$05:31:31.5 & 0.564 & 11598 & G130M, G160M & HST/ACS & F814W & 13024\\
  J095000$+$483129 & 09:50:00.7 & $+$48:31:29.4 & 0.589 & 11598 & G130M, G160M & HST/ACS & F814W & 13024\\
  PG0953$+$414 & 09:56:52.4 & $+$41:15:22.1 & 0.234 & 12038 & G130M, G160M & Pan-STARRS & $i$ & ...\\
  3C232 & 09:58:20.9 & $+$32:24:20.0 & 0.531 & 15826 & G130M & ... & ... & ...\\
  SDSSJ095914$+$320357 & 09:59:14.8 & $+$32:03:57.0 & 0.565 & 12603 & G130M & ... & ... & ...\\
  PG1001$+$291 & 10:04:02.6 & $+$28:55:35.2 & 0.329 & 12038 & G130M, G160M & HST/WFPC2 & F702W & 5949\\
  J100902$+$071343 & 10:09:02.1 & $+$07:13:43.9 & 0.456 & 11598 & G130M, G160M & HST/WFC3 & F625W & 11598\\
  RX\_J1017.5$+$4702 & 10:17:31.0 & $+$47:02:25.0 & 0.335 & 13314 & G130M & ... & ... & ...\\
  J104116$+$061016 & 10:41:17.2 & $+$06:10:16.9 & 1.27 & 12252 & G160M & HST/WFPC2 & F702W & 5984\\
  SDSSJ104335$+$115129 & 10:43:35.9 & $+$11:05:29.0 & 0.794 & 14071 & G130M & ... & ... & ...\\
  CSO295 & 10:52:05.6 & $+$36:40:40.0 & 0.609 & 14772 & G130M & ... & ... & ...\\
  RX\_J1054.2$+$3511 & 10:54:16.2 & $+$35:11:24.0 & 0.203 & 14772 & G130M & ... & ... & ...\\
  SDSSJ111443$+$525834 & 11:14:43.7 & $+$52:58:34.0 & 0.079 & 14240 & G130M & ... & ... & ...\\
  RX\_J1117.6$+$5301 & 11:17:40.5 & $+$53:01:51.0 & 0.159 & 14240 & G130M & ... & ... & ...\\
  PG1116$+$215 & 11:19:08.6 & $+$21:19:18.0 & 0.176 & 12038 & G130M, G160M & HST/WFPC2 & F606W & 5849\\
  SBS1116$+$523 & 11:19:47.9 & $+$52:05:53.0 & 0.356 & 14240 & G130M & ... & ... & ...\\
  SDSSJ112114$+$032546 & 11:21:14.0 & $+$03:25:47.0 & 0.152 & 12248 & G130M, G160M & ... & ... & ...\\
  SDSSJ112439$+$113117 & 11:24:39.4 & $+$11:31:17.0 & 0.143 & 14071 & G130M & ... & ... & ...\\
  SDSSJ112448$+$531818 & 11:24:48.3 & $+$53:18:19.0 & 0.532 & 14240 & G130M & ... & ... & ...\\
  J113327$+$032719 & 11:33:27.8 & $+$03:27:19.2 & 0.524 & 11598 & G130M, G160M & HST/ACS & F814W & 13024\\
  J113910$-$135043 & 11:39:10.7 & $-$13:50:43.6 & 0.556 & 12275 & G130M & HST/ACS & F702W & 6619\\
  PG1216$+$069 & 12:19:20.9 & $+$06:38:38.5 & 0.331 & 12025 & G130M, G160M & HST/WFPC2 & F702W & ...\\
  MRK771 & 12:32:03.6 & $+$20:09:30.0 & 0.063 & 12569 & G130M & ... & ... & ...\\
  J123304$-$003134 & 12:33:04.0 & $-$00:31:34.2 & 0.47 & 11598 & G130M, G160M & HST/ACS & F814W & 13024\\
  SDSSJ123604$+$264135 & 12:36:04.0 & $+$26:41:36.0 & 0.209 & 12248 & G130M, G160M & ... & ... & ...\\
  J124154$+$572107 & 12:41:54.0 & $+$57:21:07.4 & 0.583 & 11598 & G130M, G160M & HST/ACS & F814W & 13024\\
  PG1259$+$593 & 13:01:12.9 & $+$59:02:07.0 & 0.478 & 11541 & G130M, G160M & ... & ... & ...\\
  PG1302$-$102 & 13:05:33.0 & $-$10:33:19.0 & 0.278 & 12038 & G130M, G160M & ... & ... & ...\\
  J132222$+$464546 & 13:22:22.7 & $+$46:45:35.2 & 0.374 & 11598 & G130M, G160M & HST/ACS & F814W & 13024\\
  J134251$-$005345 & 13:42:51.6 & $-$00:53:45.3 & 0.326 & 11598 & G130M, G160M & HST/ACS & F814W & 13024\\
  QSO1500$-$4140 & 15:03:34.0 & $-$41:52:23.0 & 0.335 & 11659 & G130M & ... & ... & ...\\
  SDSSJ151237$+$012846 & 15:12:37.2 & $+$01:28:46.0 & 0.266 & 12603 & G130M & ... & ... & ...\\
  RBS1503 & 15:29:07.5 & $+$56:16:07.0 & 0.099 & 12276 & G130M & ... & ... & ...\\
  2E1530$+$1511 & 15:33:14.3 & $+$15:01:03.0 & 0.09 & 14071 & G130M & ... & ... & ...\\
  J154743$+$205216 & 15:47:43.5 & $+$20:52:16.6 & 0.264 & 13398 & G130M, G160M & HST/WFPC2 & F702W & 5099\\
  J155504$+$362847 & 15:55:04.4 & $+$36:28:48.0 & 0.714 & 11598 & G130M, G160M & HST/ACS & F814W & 13024\\
  MRK876 & 16:13:57.2 & $+$65:43:11.0 & 0.129 & 11524 & G130M & ... & ... & ...\\
  H1821$+$643 & 18:21:57.2 & $+$64:20:36.2 & 0.297 & 12038 & G130M, G160M & HST/ACS, Pan-STARRS
 & F814W,$i$ & 13024\\
  J213135$-$120704 & 21:31:35.3 & $-$12:07:04.8 & 0.501 & 13398 & G160M & HST/WFPC2 & F702W & 5143\\
  J213745$-$143255 & 21:37:45.2 & $-$14:32:55.8 & 0.2 & 13398 & G130M, G160M & HST/WFPC2 & F702W & 5343\\
  RBS1768 & 21:38:49.9 & $-$38:28:40.0 & 0.183 & 12936 & G130M, G160M & ... & ... & ...\\
  PHL1811 & 21:55:01.5 & $-$09:22:25.0 & 0.19 & 12038 & G130M, G160M & Pan-STARRS
 & $i$ & ...\\
  J225357$+$160853 & 22:53:57.7 & $+$16:08:53.6 & 0.859 & 13398 & G130M, G160M & HST/WFPC2 & F702W & 6619\\
  MRC2251$-$178 & 22:54:05.9 & $-$17:34:55.0 & 0.066 & 12029 & G130M, G160M & ... & ... & ...\\
  RBS2000 & 23:24:44.7 & $-$40:40:49.0 & 0.174 & 13448 & G130M, G160M & ... & ... & ...\\
\hline\end{tabular}
	\label{QSOs}
	
\end{table*}

 \section{Absorption properties}
In Table \ref{absorption}, we present the measured properties of CGM {\HI} absorption studied in this work. The absorbers are detected in the spectrum of the background quasar in each field presented in the first column of this table. The host galaxies and absorption redshifts can be found in the second and third columns, respectively. We present the rest-frame equivalent width of {\Lya} absorption, $W_{r}(1215)$, and its column density in columns 4 and 5, respectively. In column 4, there are five systems that {\Lyb} is used to measure their EW co-rotation fraction as the {\Lya} is not covered by the QSO spectra (see table note). {\fewcorot} is presented in column 7 and the last column lists the source of the {\HI} column density measurement.  
 
\begin{table*}
\centering
\renewcommand{\arraystretch}{1.1}
\caption{Absorption properties}
\label{absorption}
\begin{threeparttable}

\begin{tabular}{l l c c c c c }
\hline
\hline
\\
  \multicolumn{1}{l}{Quasar} &
  \multicolumn{1}{l}{Galaxy} &
  \multicolumn{1}{c}{$z_{\rm abs}$} &
  \multicolumn{1}{c}{$W_{r}(1215)$~({\AA})} &
  \multicolumn{1}{c}{\colden} &
  \multicolumn{1}{c}{\fewcorot\tnote{b}} &
  \multicolumn{1}{c}{References\tnote{c}}\\
\hline
  MRK335 & NGC7817 & 0.006936 & 0.424 $\pm$0.028 & 13.32 $\pm$0.03  & 0.15 & 1 \\
  J035128$-$142908 & J0351G1 & 0.356706 & 1.198 $\pm$0.018 & 16.86 $\pm$0.03  & 0.69 & 2\\
  PKS0405$-$123 & PKS0405G1 & 0.360814 & 0.795 $\pm$0.007 & 15.26 $\pm$0.06 & 0.70 & 1\\
  J040748$-$121136 & J0407G1 & 0.495111 & 0.141 $\pm$0.005\tnote{a} & 14.34 $\pm$0.56  & 0.57 & 2\\
  J045608$-$215909 & J0456G1 & 0.381664 & 0.60 $\pm$0.02 & 15.10 $\pm$0.39  & 0.78 & 2\\
  J045608$-$215909 & J0456G2 & 0.27799 & 0.752 $\pm$0.013 & 14.78 $\pm$0.22 & 0.60 & 3\\
  PG0804$+$761 & UGC04238 & 0.005118 & 0.082 $\pm$0.004 & 12.6 $\pm$0.08 & 0.46 & 1\\
  J085334$+$434902 & J0853G1 & 0.090763 & 0.577 $\pm$0.009 & 14.53 $\pm$0.04  & 0.37 & 4\\
  J085334$+$434902 & J0853G2 & 0.163718 & 5.764 $\pm$0.053 & 19.93 $\pm$0.01 & 0.57 & 2\\
  SDSSJ091052$+$333008 & NGC2770 & 0.006115 & 0.363 $\pm$0.061 & 13.22 $\pm$0.11 & 0.07 & 1\\
  TON1015 & NGC2770 & 0.006064 & 0.334 $\pm$0.033 & 13.24 $\pm$0.08 & 0.15 & 1\\
  TON1009 & NGC2770 & 0.006597 & 0.338 $\pm$0.023 & 13.38 $\pm$0.12  & 0.29 & 1\\
  FBQSJ0908$+$3246 & NGC2770 & 0.006467 & 0.345 $\pm$0.056 & 13.81 $\pm$0.05 & 0.58 & 1\\
  SDSSJ091127$+$325337 & NGC2770 & 0.00688 & 0.243 $\pm$0.047 & 14.00 $\pm$0.20 & 0.00 & 1\\
  J091440$+$282330 & J0914G1 & 0.244096 & 0.791 $\pm$0.021 & 15.55 $\pm$0.03 & 0.23 & 2\\
  J094331$+$053131 & J0943G1 & 0.354399 & 2.413 $\pm$0.076 & 16.46 $\pm$0.03 & 0.94 & 2\\
  J094331.61$+$053131.4 & J0943G2 & 0.548808 & 0.123 $\pm$0.025\tnote{a} & 14.61 $\pm$0.08 & 1.00 & 4\\
  J095000$+$483129 & J0950G1 & 0.211585 & 1.297 $\pm$0.017 & 18.48 $\pm$0.19 & 0.29 & 3\\
  PG0953$+$414 & PG0953G1 & 0.058755 & 0.27 $\pm$0.01 & 13.96 $\pm$0.07 & 0.68 & 5\\
  SDSSJ095914$+$320357 & NGC3067 & 0.004987 & 0.556 $\pm$0.024 & 16.23 $\pm$1.43 & 0.69 & 1\\
  3C232 & NGC3067 & 0.004805 & 6.63 $\pm$0.09 & 20.09 $\pm$0.02 & 0.52 & 1\\
  PG1001$+$291 & PG1001G1 & 0.137458 & 0.717 $\pm$0.008 & 14.98 $\pm$0.03 & 0.64 & 4\\
  J100902$+$071343 & J1009G1 & 0.227858 & 0.98 $\pm$0.02 & 17.23 $\pm$0.16 & 0.52 & 3\\
  RX\_J1017.5$+$4702 & NGC3198 & 0.002079 & 0.057 $\pm$0.019 & 13.18 $\pm$0.12 & 1.00 & 1\\
  J104116$+$061016 & J1041G1 & 0.441546 & 1.146 $\pm$0.027 & 18.19 $\pm$0.14 & 0.94 & 3\\
  SDSSJ104335$+$115129 & NGC335 & 0.002341 & 0.762 $\pm$0.068 & 14.53 $\pm$0.12 & 0.85 & 1\\
  RX\_J1054.2$+$3511 & NGC3432 & 0.002222 & 0.234 $\pm$0.072 & 13.58 $\pm$0.12 & 0.09 & 1\\
  CSO295 & NGC3432 & 0.002204 & 0.963 $\pm$0.063 & 15.05 $\pm$0.37 & 0.65 & 1\\
  PG1116$+$215 & PG1116G1 & 0.138513 & 0.516 $\pm$0.004 & 16.20 $\pm$0.03 & 0.59 & 5\\
  PG1116$+$215 & PG1116G2 & 0.166152 & 0.780 $\pm$0.004 & 14.71 $\pm$0.05 & 0.74 & 5\\
  RX\_J1121.2$+$0326 & NGC3633 & 0.008934 & 0.19 $\pm$0.08 & 13.70 $\pm$0.18 & 0.00 & 1\\
  RX\_J1117.6$+$5301 & NGC3631 & 0.003763 & 0.447 $\pm$0.038 & 13.17 $\pm$0.10 & 0.30 & 1\\
  SDSSJ112448$+$531818 & NGC3631 & 0.003701 & 0.241 $\pm$0.052 & 13.18 $\pm$0.11 & 0.75 & 1\\
  SDSSJ111443$+$525834 & NGC3631 & 0.003837 & 0.160 $\pm$0.064 & 13.52 $\pm$0.09  & 0.36 & 1\\
  SDSSJ112114$+$032546 & CGCG039137 & 0.023493 & 0.50 $\pm$0.09 & 14.27 $\pm$0.06 & 1.00 & 1\\
  SDSSJ112439$+$113117 & NGC3666 & 0.003487 & 0.664 $\pm$0.044 & 15.53 $\pm$0.67 & 0.61 & 1\\
  SDSSJ112448$+$531818 & UGC06446 & 0.002202 & 0.261 $\pm$0.051 & 14.07 $\pm$0.04 & 0.65 & 1\\
  J113327$+$032719 & J1133G1 & 0.154198 & 0.686 $\pm$0.024 & 16.76 $\pm$0.96  & 1.00 & 3\\
  J113910$-$135043 & J1139G1 & 0.219799 & 0.099 $\pm$0.008\tnote{a} & 14.20 $\pm$0.07 & 0.63 & 2\\
  J113910$-$135043 & J1139G2 & 0.212036 & 0.268 $\pm$0.006\tnote{a} & 15.33 $\pm$0.04 & 0.03 & 2\\
  J113910$-$135043 & J1139G3 & 0.319419 & 0.625 $\pm$0.008\tnote{a} & 16.19 $\pm$0.03 & 0.46 & 2\\
  J113910$-$135043 & J1139G4 & 0.204418 & 1.26 $\pm$0.02 & 16.28 $\pm$0.34 & 0.69 & 3\\
  PG1216$+$069 & PG1216G1 & 0.124006 & 1.417 $\pm$0.008 & [16.06,19] & 0.17 & 5\\
  MRK771 & NGC4529 & 0.00849 & 0.229 $\pm$0.012 & 13.03 $\pm$0.49 & 0.39 & 1\\
  J123304$-$003134 & J1233G1 & 0.318659 & 0.964 $\pm$0.024 & 15.72 $\pm$0.02 & 0.43 & 2\\
  SDSSJ123604$+$264135 & NGC4565 & 0.003897 & 0.348 $\pm$0.032 & 13.31 $\pm$0.14 & 0.17 & 1\\
  J124154$+$572107 & J1241G1 & 0.205584 & 1.071 $\pm$0.012 & 18.38 $\pm$0.16 & 0.81 & 3\\
  J124154$+$572107 & J1241G2 & 0.218094 & 0.750 $\pm$0.016 & 15.59 $\pm$0.12 & 0.27 & 2\\
  PG1259$+$593 & UGC08146 & 0.002274 & 0.244 $\pm$0.009 & 13.04 $\pm$0.14 & 0.59 & 1\\
  PG1302$-$102 & NGC4939 & 0.011482 & 0.09 $\pm$0.01 & 13.23 $\pm$0.04 & 0.00 & 1\\
  J132222$+$464546 & J1322G1 & 0.214527 & 1.103 $\pm$0.022 & 17.49 $\pm$0.2 & 0.58 & 3\\
  J134251$-$005345 & J1342G1 & 0.227256 & 1.891 $\pm$0.033 & 18.83 $\pm$0.05 & 0.39 & 2\\
  QSO1500$-$4140 & NGC5786 & 0.010422 & 0.16 $\pm$0.04 & 13.85 $\pm$0.08 & 1.00 & 1\\
  SDSSJ151237$+$012846 & UGC09760 & 0.006804 & 0.44 $\pm$0.07 & 14.50 $\pm$0.15 & 0.10 & 1\\
  2E1530$+$1511 & NGC5951 & 0.006046 & 0.646 $\pm$0.054 & 13.73 $\pm$0.05 & 0.65 & 1\\
  J154743$+$205216 & J1547G1 & 0.096155 & 0.228 $\pm$0.013 & 13.75 $\pm$0.03 & 0.04 & 2\\
  
\hline\end{tabular}

\end{threeparttable}
\end{table*}

\begin{table*}
\addtocounter{table}{-1}
\centering
\renewcommand{\arraystretch}{1.1}
\caption{Absorption Properties continued}
\begin{threeparttable}

\begin{tabular}{l l c c c  c c }
\hline
\hline
\\
  \multicolumn{1}{l}{Quasar} &
  \multicolumn{1}{l}{Galaxy} &
  \multicolumn{1}{c}{$z_{\rm abs}$} &
  \multicolumn{1}{c}{$W_{r}(1215)$~({\AA})} &
  \multicolumn{1}{c}{\colden} &
  \multicolumn{1}{c}{\fewcorot\tnote{b}} &
  \multicolumn{1}{c}{References\tnote{c}}\\
\hline
  J155504$+$362847 & J1555G1 & 0.189054 & 0.977 $\pm$0.084 & 17.52 $\pm$0.22 & 0.64 & 3\\
  MRK876 & NGC6140 & 0.00311 & 0.388 $\pm$0.005 & 13.49 $\pm$0.15 & 0.63 & 1\\
  H1821$+$643 & H1821G1 & 0.224874 & 1.03 $\pm$0.02 & 15.55 $\pm$0.02  & 0.32 & 5\\
  J213135$-$120704 & J2131G1 & 0.429825 & 3.189 $\pm$0.038 & 19.88 $\pm$0.10 & 0.58 & 2\\
  RBS1768 & ESO343G014 & 0.031304 & 0.51 $\pm$0.01 & 13.05 $\pm$0.08  & 0.00 & 1\\
  J213745$-$143255 & J2137G1 & 0.07532 & 0.279 $\pm$0.007 & 13.96 $\pm$0.02 & 0.86 & 2\\
  PHL1811 & PHL1811G2 & 0.323091 & 0.20 $\pm$0.01 & 13.61 $\pm$0.03  & 0.00 & 4\\
  PHL1811 & PHL1811G1 & 0.176514 & 0.470 $\pm$0.003 & 14.93 $\pm$0.03  & 0.98 & 5\\
  PHL1811 & PHL1811G3 & 0.157814 & 0.153 $\pm$0.004 & 13.26 $\pm$0.09 & 0.83 & 5\\
  MRC2251$-$178 & MCG0358009 & 0.030114 & 0.066 $\pm$0.005 & 13.08 $\pm$0.04 & 0.59 & 1\\
  J225357$+$160853 & J2253G1 & 0.153766 & 0.937 $\pm$0.022 & 16.04 $\pm$0.73 & 0.45 & 3\\
  J225357$+$160853 & J2253G2 & 0.352607 & 0.766 $\pm$0.029 & 14.53 $\pm$0.05 & 0.74 & 2\\
  J225357$+$160853 & J2253G3 & 0.390642 & 0.934 $\pm$0.043 & 15.19 $\pm$0.04 & 0.01 & 4\\
  RBS2000 & IC5325 & 0.005356 & 0.045 $\pm$0.013 & 12.85 $\pm$0.10 & 0.00 & 1\\
\hline\end{tabular}
	\begin{tablenotes}
            \item[a] The rest-frame {\Lyb} equivalent width is reported because the {\Lya} was not covered by the background QSO spectra. In these absorption systems the {\Lyb} is used for the purpose of measuring the co-rotation fraction.
            \item[b] Uncertainties range between $0.001-0.008$ as determined from a bootstrap analysis where we varied the galaxy and absorption redshifts within their error bars.
             \item[c] ~{\HI} absorption column density  reference: (1) \cite{French2020}, (2) \cite{pointon19}, (3) \cite{sameer2024}, (4) this work, (5) \cite{Tripp2008}.
        \end{tablenotes}
        \end{threeparttable}
\end{table*}

\newpage
\bsp	
\label{lastpage}
\end{document}